\documentclass[10pt,preprint]{aastex}
\usepackage{graphicx}% Include figure files
\usepackage{dcolumn}% Align table columns on decimal point
\usepackage{bm}% bold math
\usepackage{epsfig,amsmath}
\usepackage[usenames,dvipsnames]{color}
\usepackage[pagebackref=false, colorlinks=true]{hyperref}
\usepackage{appendix}
\usepackage{textcomp}
\usepackage{subfigure}
\definecolor{redish}{rgb}{0.7,0.2,0.0}  % color defined in (r=red,g=green,b=blue) model
\definecolor{bluish}{rgb}{0.2,0.5,0.8}

\hypersetup{linkcolor=redish,          % color of internal links
                  citecolor=blue,        % color of links to bibliography
                  filecolor=magenta,      % color of file links
                  urlcolor=bluish}          % color of the url

\DeclareFontFamily{U}{rsfs}{}         % Formal Script            %
\DeclareFontShape{U}{rsfs}{m}{n}{<5> rsfs5 <6><7> rsfs7          %
  <8><9><10><10.95><12><14.4><17.28><20.74><24.88> rsfs10}{}     %
\DeclareMathAlphabet{\mathfs}{U}{rsfs}{m}{n}                     %

\newcommand{\ba}{\nopagebreak[3]\begin{eqnarray}}
\newcommand{\ea}{\end{eqnarray}}

\newcommand{\f}{\frac}

\def \o{\omega}

\def \a{\alpha}

\def \O{\Omega}
\def \g{\gamma}
\def \s{\sigma}
\def \e{\epsilon}
\def \p{\partial}

\def \th{\theta}
\def \r{\rho}
\def \ve{\varepsilon}
\def \ti{\tilde}
\begin{document}
\title{Dragging of inertial frames inside the rotating neutron stars}
\author{Chandrachur Chakraborty,
Kamakshya Prasad Modak, Debades Bandyopadhyay}
\affil{Astroparticle Physics and Cosmology Division \\
Saha Institute of Nuclear Physics, Kolkata 700064, India}

\email{chandrachur.chakraborty@saha.ac.in}
\email{kamakshya.modak@saha.ac.in}

\begin{abstract}
We derive the exact frame-dragging rate inside rotating neutron
stars. This formula is applied to show that the frame-dragging rate 
monotonically decreases from the centre to the 
surface of the neutron star along the pole. In case of 
frame-dragging
rate along the equatorial distance, it decreases initially away from the 
centre, becomes negligibly small well before  the surface of the neutron
star, rises 
again and finally approaches to a small value at the surface. The appearance
of local maximum and minimum in this case is the result of the dependence of 
frame-dragging frequency on the distance and angle. 
Moving from the equator to the pole, it is observed that this local maximum and
minimum in the frame-dragging rate along the equator
disappears after crossing a critical angle. 
It is also noted that the positions 
of local maximum and minimum of the frame-dragging rate along the
equator depend on the rotation frequency and central energy density of a 
particular pulsar. 
\end{abstract}

\maketitle

\section{Introduction} 
Compact astrophysical objects such as neutron stars and black holes are the
laboratories for the study of Einstein's general relativity in strong 
gravitational fields. The frame-dragging is one such important general 
relativistic effect as demonstrated by Lense and Thirring \cite{LT}.  
A stationary spacetime with angular momentum 
shows an effect by which the locally inertial frames are dragged
along the rotating spacetime. This makes any test gyroscope in such spacetime
precess with a certain frequency called the frame-dragging frequency or
the Lense-Thirring (LT) precession frequency $(\O_{LT})$.
The Lense-Thirring frequency is 
proportional to the angular momentum and compactness of the rotating
astrophysical compact object. This effect for 
a test gyroscope had been calculated and was
shown to fall with the inverse cube of the distance of
the test gyroscope from the source and vanishes at large enough distances
where the curvature effects are small. 
 The precession frequency is thus expected to be
larger near the surface of a neutron star and 
in its interior, rather than at large distance from the star.
 
The precise mass measurement of PSR J0348+0432 confirmed the existence  
of a massive neutron star ( $> 2 M_{\odot}$) \citep{anto}. It is also known 
that some of them are observed to possess very high angular velocities.
Hence the spacetime curvature would be much higher in the surroundings of 
those massive neutron stars and the frame dragging effect also
becomes very significant in 
the strong gravitational fields of those rotating neutron stars. 
It should be noted that the inertial frames 
are dragged not only outside but also inside the rotating neutron stars.

The theoretical prescription
to determine the rate of the frame-dragging precession inside
the rotating neutron star was first given by Hartle \cite{hr67}.
In this formalism, one can estimate
the frame-dragging precession rate inside a slowly rotating 
($\O R<<c$, where $R$ is the radius of the pulsar, $c$ is
the speed of the light in vacuum) neutron star.
The final expression of frame-dragging precession rate depends solely on $r$, 
the distance from the centre of the star,  due to the slow rotation 
approximation, in Hartle's formalism. 
It was observed that the frame-dragging frequency 
was higher at the centre of the star than the frame dragging 
frequency at the surface. 
The maximum frame dragging frequency at the centre $(r=0)$
would never exceed the frequency of the rotating neutron star.
The frame-dragging effect was applied
to various astrophysical problems using Hartle's formalism. 
Hartle studied this effect on the equilibrium structures of rotating neutron 
stars \citep{hr68}. The impact of frame dragging on the Kepler frequency was
investigated by Glendenning and Weber \citep{GW}. It was also demonstrated how 
this effect might influence
the moment of inertia of a rotating neutron star \citep{web}. Furthermore, 
Morsink and Stella studied the role of frame dragging in explaining the 
Quasi Periodic Oscillations of accreting neutron stars \citep{mor}.
 Morsink and Stella estimated the 
precession frequency $\nu_p$ of the disk's orbital plane 
about the star's axis of symmetry as the difference between
the frequency of oscillations of the particle along the 
longitude and latitude $(2\pi\nu_p=d\phi/dt-d\theta/dt)$
observed at infinity. This expression contains
the total precession frequency of the disk's orbital plane
due to the Lense-Thirring (LT) effect as well as the star's oblateness.
Their calculation introduced zero angular momentum observer (ZAMO) and 
the precession frequency $\nu_p$ was observed at infinity. It was found that
the LT frequency was proportional to the ZAMO frequency on the equatorial plane
in the slow rotation limit. This is similar to the
Hartle's formalism\citep{hr67} where the angular
velocity $(d\phi/dt)$ acquired by an observer who falls
freely from infinity to the point $(r,\th)$, is taken as {\it the
rate of rotation of the inertial frame at that point relative to
the distant stars.} 

In this present manuscript we derive the exact Lense-Thirring
precession frequency $(\Omega_{LT})$ which is measured by a Copernican observer
of a gyroscope such as the Gravity Probe B satellite in a 
realistic orbit\cite{jh}. In this case, $\Omega_{LT}$ would not only be the 
function of $\omega$ but a complicated function of other
metric components also even in the slow rotation limit.

We should note that inside the rapidly rotating stars, one should not a 
priori expect the similar variation of the precession rates along the
equatorial and polar plane. 
Thus the frame-dragging frequency should depend also on the colatitude ($\th$)
of the position of the test gyroscope. This did not arise in the formalism of
Hartle due to the slow-rotation limit. 
We also note that the LT precession must depend
on both the radial distance ($r$) and the colatitude ($\th$) (see Eq.(14.34)
of ~\cite{jh}) in very weak gravitational fields (far away 
from the surface of the rotating object). 

The exact Lense-Thirring precession rate in strongly
curved stationary spacetime had been discussed in detail by Chakraborty
and Majumdar~\cite{cm}. Later, Chakraborty and Pradhan~\cite{cp}
applied this formulation in various stationary and 
axisymmetric spacetimes.
Our main motivation of this paper is to compute the exact LT precession rate
inside the rotating neutron star.
In this article we avoid all types of approximations 
and assumptions to obtain the exact LT precession rate inside the rotating
neutron stars. 

The paper is organized as follows. 
In section 2 we present the basic equations of frame-dragging
effect inside the rotating neutron stars. 
The numerical method, which 
has been adopted in the whole paper, is discussed in section 3.
We discuss our results in section 4. Finally we conclude
in section 5 with a summary.

\section{Basic equations of frame-dragging effect inside the rotating neutron stars}
The rotating equilibrium models considered in this 
paper are stationary and axisymmetric. Thus we can write the metric
inside the rotating neutron star as the following Komatsu-Eriguchi-Hachisu
(KEH)~\cite{keh} form:
\begin{equation}
 ds^2=-e^{\g+\s}dt^2+e^{2\a}(dr^2+r^2d\th^2)+e^{\g-\s}r^2\sin^2\th(d\phi-\o dt)^2
\label{met}
\end{equation}
where $\g,\,\s,\,\a,\,\o$ are the functions of $r$ and $\th$ only.
In the whole paper we have used the geometrized unit ($G=c=1$).
We assume that the matter source is a perfect fluid with a
stress-energy tensor given by
\begin{equation}
 T^{\mu\nu}=(\r_0+\r_i+P)u^\mu u^\nu+Pg^{\mu\nu}
\end{equation}
where $\r_0$ is the rest energy density, $\r_i$ is the
internal energy density, $P$ is the pressure and $u^\mu$
is the matter four velocity. We are further assuming that
there is no meridional circulation of the matter so that
the four-velocity $u^\mu$ is simply a linear combination
of time and angular Killing vectors. Now, we have to
calculate the frame-dragging rate based on the above metric
and this will gives us the exact frame-dragging rate inside a
rotating neutron star.

We know that the vector field  corresponding to the LT precession
co-vector can be expressed as
\begin{eqnarray}
\O_{LT}=\f{1}{2} \f{\e_{ijl}}{\sqrt {-g}} \left[g_{0i,j}\left(\p_l -
\f{g_{0l}}{g_{00}}\p_0\right) -\f{g_{0i}}{g_{00}} g_{00,j}\p_l\right]
\label{s25}
\end{eqnarray}
For the axisymmetric spacetime, the only
non-vanishing component is $g_{0i} = g_{0\phi}$,
$i = \phi$ and $j, l = r, \theta$; substituting these in
Eq. (\ref{s25}), the LT precession frequency vector is obtained as:
\begin{eqnarray}
 \O_{LT}=\f{1}{2\sqrt {-g}} \left[\left(g_{0\phi,r}-\f{g_{0\phi}}{g_{00}}
g_{00,r}\right)\p_{\theta}
 - \left(g_{0\phi,\theta}- \f{g_{0\phi}}{g_{00}} g_{00,\theta}\right)\p_r\right]
\label{k3}
\end{eqnarray}
As the above expression has been expressed in the co-ordinate basis
we have to convert it into the orthonormal basis. Thus, 
 in the orthonormal basis, with our choice of polar co-ordinates, $\O_{LT}$ 
can be written as\cite{cm}
\begin{equation}
 \vec{\O}_{LT}=\f{1}{2\sqrt {-g}}\left[\sqrt{g_{rr}}\left(g_{0\phi,\th}
-\f{g_{0\phi}}{g_{00}} g_{00,\th}\right)\hat{r}
+\sqrt{g_{\th\th}}\left(g_{0\phi,r}-\f{g_{0\phi}}{g_{00}}
g_{00,r}\right)\hat{\th}\right]
\label{ltp}
\end{equation}
where $\hat{\th}$ is the unit vector along the direction $\theta$
and $\hat{r}$ is the unit vector along the direction $r$.
We note that the above formulation is valid only in the timelike
spacetimes, not in the lightlike or spacelike regions.

Now, we can apply the above Eq.(\ref{ltp}) to determine
the exact frame-dragging rate inside the rotating neutron star
of which the metric could be determined from the line-element(\ref{met}).
The various metric components can be read off from the metric.
Likewise,
\begin{equation}
 \sqrt{-g}=r^2e^{2\a+\g}\sin\th 
\end{equation}

In orthonormal coordinate basis, the exact Lense-Thirring 
precession rate inside the rotating neutron star is:
\begin{eqnarray}\nonumber
&& \vec{\O}_{LT}=\f{e^{-(\a+\s)}}{2(\o^2r^2\sin^2\th-e^{2\s})}.
\label{ltnp}
\\ \nonumber
&& \left[\sin\th[r^3\o^2\o_{,r}\sin^2\th+e^{2\s}(2\o+r\o,_r-2\o r \s,_r)]\hat{\th}
+[r^2\o^2\o_{,\th}\sin^3\th+e^{2\s}(2\o \cos\th+\o_{,\th}\sin\th
-2\o\s_{,\th}\sin\th)]\hat{r}\right]
\\
\end{eqnarray}
and the modulus of the above LT precession rate is
\begin{eqnarray}\nonumber
&&\O_{LT}=|\vec{\O}_{LT}(r,\th)|=\f{e^{-(\a+\s)}}{2(\o^2r^2\sin^2\th-e^{2\s})}.
\label{ltn}
\\  \nonumber           
&&\left[\sin^2\th[r^3\o^2\o_{,r}\sin^2\th+e^{2\s}(2\o+r\o,_r-2\o r \s,_r)]^2
+[r^2\o^2\o_{,\th}\sin^3\th+e^{2\s}(2\o \cos\th+\o_{,\th}\sin\th
-2\o\s_{,\th}\sin\th)]^2\right]^{\f{1}{2}}
\\
\end{eqnarray}
 As a vector quantity the expression of $\vec{\O}_{LT}$ (Eq.\ref{ltnp})
depends on the coordinate frame (i.e. polar coordinates 
$(r,\th)$ which is used here) but the modulus
of $\vec{\O}_{LT}$ or $|\vec{\O}_{LT}|$ (Eq.\ref{ltn}) must be coordinate
frame independent. Here, we use only the modulus of $\O_{LT}$ 
in the rest of our manuscript.  

 To calculate the frame-dragging precession frequency at
the centre of the neutron star we substitute $r=0$ in 
Eq. (\ref{ltn}) and we obtain
\begin{eqnarray}
\O_{LT}|_{r=0}=\left[\f{{e^{-(\a+\s)}}}{2}\left[4\o^2+(\o_{,\th}-2\o\s_{,\th})^2\sin^2\th
+4\o\cos\th \sin\th(\o_{,\th}-2\o\s_{,\th})\right]^{\f{1}{2}}\right]_{|{r=0}}
\end{eqnarray}
solving numerically $\o_{,\th}$ and $\s_{,\th}$ at the centre
we get the value zero for both of them.
 Thus, we obtain the frame-dragging precession rate 
\begin{eqnarray}
\O_{LT}|_{r=0}=\o{e^{-(\a+\s)}}|_{r=0}
\end{eqnarray}
at the center ($r=0$) of a rotating neutron star.

Following KEH we can write the general relativistic field
equations determining $\s, \g$ and $\o$ as
\begin{eqnarray}
\Delta\left[\s e^{\f{\g}{2}}\right]&=&S_{\s}(r,\mu)
\label{dr}
\\
\left(\Delta+\f{1}{r}\p_r-\f{\mu}{r^2}\p_{\mu}\right)
\left[\g e^{\f{\g}{2}}\right]&=&S_{\g}(r,\mu)
\label{dg}
\\
\left(\Delta+\f{2}{r}\p_r-\f{2\mu}{r^2}\p_{\mu}\right)\left[\o e^{\f{\g}{2}-\s}\right]
&=&S_{\o}(r,\mu) 
\label{do}
\end{eqnarray}
where 
\begin{eqnarray}
\Delta \equiv \p^2_{r}+\f{2}{r}\p_{r} +
\f{1-\mu^2}{r^2}\p^2_{\mu}-\f{2\mu}{r^2}\p_{\mu}+\f{1}{r^2(1-\mu^2)}\p^2_{\phi}
\end{eqnarray}
is the flat-space spherical coordinate Laplacian,
$\mu=\cos\th$ and $S_{\s},S_{\g},S_{\o}$ are the effective source terms
that include the nonlinear and coupling terms. The effective source terms
are given by

\begin{eqnarray}
S_{\s}(r,\mu) &=& e^{\g/2}\left\{8\pi (\r_0+\r_i 
+ P)e^{2\a} \f{1+v^2}{1-v^2}+ r^2
(1-\mu^2)e^{-2\s}\left[\o_{,r}^2  +
\f{1-\mu^2}{r^2}\o_{,\mu}^2\right] +
\f{1}{r}\g_{,r} - \f{\mu}{r^2}\g_{,\mu}
 \right. \nonumber  \\ && \nonumber \\
 &&\left.  + \f{\s}{2}\left[16\pi Pe^{2\a} 
- \f{1}{r}\g_{,r} + \f{\mu}{r^2}\g_{,\mu} -
 \f{1}{2}\g_{,r}^2  - \f{1-\mu^2}{2r^2} 
\g_{,\mu}^2 \right] \right\}, \label{eq:DiffEq_sigma}
\end{eqnarray}

\begin{eqnarray}
S_{\g}(r,\mu)&=&e^{\g/2}\left[16\pi e^{2\a}P +
\f{\g}{2}\left(16\pi e^{2\a}P-\f{1}{2}\g_{,r}^2 -
 \f{1-\mu^2}{2r^2} \g_{,\mu}^2\right)\right], 
\label{diffeq_g}
\end{eqnarray}

\begin{eqnarray}\nonumber
S_{\o}(r,\mu)&=&e^{\g/2 - \s} \left\{- 16\pi
\f{(\r_0+\r_i  + P)(\O - \o)}{1- v^2}e^{2\a}
+\o\left[-8\pi \f{(\r_0+\r_i)(1+ v^2) +
2Pv^2}{1- v^2} e^{2\a}-\f{1}{r}\left(\f{1}{2}\g_{,r} + 2\s_{,r}\right) \right. \right.
\\
&+&\left. \left. \f{\mu}{r^2}\left(\f{1}{2}\g_{,\mu} + 2\s_{,\mu}\right)
+\s_{,r}^2 -\f{1}{4}\g_{,r}^2 +\f{1-\mu^2}{4r^2}
(\g_{,\mu}^2+4\s_{,\mu}^2)- r^2(1-\mu^2)e^{-2\s}
\left(\o_{,r}^2 + \f{1-\mu^2}{r^2}
\o_{,\mu}^2\right)\right]\right\}.
 \label{eq:DiffEq_omega}
\end{eqnarray}
where $\O$ is the angular velocity of the matter as measured at
infinity and $v$ is the proper velocity of the matter with respect
to a zero angular momentum observer. The proper velocity of the
matter is given by 
\begin{eqnarray}
 v=(\O-\o)re^{-\s}\sin\th
\end{eqnarray}
and the coordinate components of the four-velocity of the
matter can be written as
\begin{eqnarray}
 u^{\mu}=\f{e^{-(\s+\g)/2}}{\sqrt{1-v^2}}[1,0,0,\O]
\end{eqnarray}

Following Cook, Shapiro, Teukolsky \cite{cst} we can write
the another field equation which determines $\a$ and is given by
\begin{eqnarray}\nonumber
&&\a_{,\mu} = -\frac{1}{2}(\g_{,\mu} + \s_{,\mu}) - 
\left\{(1-\mu^2)(1+r\g_{,r})^2 + [\mu -(1-\mu^2)\g_{,\mu}]^2 \right\}^{-1} \\ \notag \\
&&\left\{\frac{1}{2}\left[r^2(\g_{,rr} + \g_{,r}^2) - (1-\mu^2)(\g_{,\mu}^2
 +\g_{,\mu\mu})\right][-\mu + (1-\mu^2)\g_{,\mu}]+\f{3}{2}\mu\g_{,\mu}[-\mu + (1-\mu^2)\g_{,\mu}]
 \right. \notag \\ \notag \\
&& +\frac{1}{4}[-\mu + (1-\mu^2)\g_{,\mu}]\left[r^2(\g_{,r} + \s_{,r})^2 
- (1-\mu^2)(\g_{,\mu} + \s_{,\mu})^2\right]+  r \g_{,r} \left[\f{\mu}{2}+\mu r\g_{,r}
+\f{(1-\mu^2)\g_{,\mu}}{2}\right]\notag \\ \notag \\
&&
- (1-\mu^2)r(1+r\g_{,r})\left[\g_{,r\mu}
 + \g_{,\mu}\g_{,r} + \frac{1}{2}(\g_{,\mu} + \s_{,\mu})(\g_{,r} + \s_{,r}) 
\right]+\frac{1}{4}(1-\mu^2)e^{-2\s}. \notag \\ \notag \\
&&\left. \left[r^4\mu\o_{,r}^2
+(1-\mu^2)[2r^3\o_{,r}\o_{,\mu}-\mu r^2\o_{,\mu}^2
+2r^4\g_{,r}\o_{,r}\o_{,\mu}]
-r^2(1-\mu^2)\g_{,\mu}[r^2\o_{,r}^2-
(1-\mu^2)\o_{,\mu}^2]\right]\right\}
\label{eq:DiffEq_alpha}
\end{eqnarray}

\section{Numerical method}
Here we adopt the rotating neutron star ({\tt rns}) code
based on the KEH \cite{keh} method and
written by Stergioulas \cite{str} to obtain the 
frame-dragging rate inside the rotating neutron stars.
The equations for the gravitational and matter fields were solved
on a discrete grid using a combination of integral and finite 
difference techniques. The computational domain of the problem is
$0\leq r\leq\infty$ and $0\leq\mu\leq1$. It is easy to deal
with finite radius rather than the infinite domain with via a
coordinate transformation to a new radial coordinate $s$ which covers
the infinite radial span in a finite coordinate interval $0\leq s\leq1$.
This new radial coordinate $s$ is defined by    
\begin{eqnarray}
 r=r_e.\f{s}{1-s}
\label{rs}
\end{eqnarray}
Thus, $s=\f{1}{2}$ represents the radius of the equator ($r_e$) of the pulsar
and $s=1$ represents the infinity.
\\
The three elliptical field equations (\ref{dr})-(\ref{do}) were solved 
by an integral Green's function approach following the KEH. Taking into
account the equatorial and axial symmetry in the configurations we 
can find the three metric coefficients $\s, \g, \o$ which can be written as
\begin{eqnarray}\nonumber
 \s(s,\mu)&=&-e^{-\f{\g}{2}}\sum\limits_{n=0}^{\infty} P_{2n}(\mu)
\left[\left(\f{1-s}{s}\right)^{2n+1}\int_0^s\f{s'^{2n}ds'}{(1-s')^{2n+2}}
\int_0^1d\mu'P_{2n}(\mu')\bar{S}_{\s}(s',\mu') \right. 
\\
&+&\left. \left(\f{s}{1-s}\right)^{2n}\int_s^1\f{(1-s')^{2n-1}ds'}{s'^{2n+1}}
\int_0^1d\mu'P_{2n}(\mu')\bar{S}_{\s}(s',\mu')\right] ,
\label{r}
\\ \nonumber
 \g(s,\mu)&=&-\f{2e^{-\f{\g}{2}}}{\pi}\sum\limits_{n=1}^{\infty} \f{\sin[(2n-1)\th]}{(2n-1)\sin\th}
\left[\left(\f{1-s}{s}\right)^{2n}\int_0^s\f{s'^{2n-1}ds'}{(1-s')^{2n+1}}
\int_0^1d\mu'\sin[(2n-1)\th']\bar{S}_{\g}(s',\mu') \right. 
\\
&+&\left. \left(\f{s}{1-s}\right)^{2n-2}\int_s^1\f{(1-s')^{2n-3}ds'}{s'^{2n-1}}
\int_0^1d\mu'\sin[(2n-1)\th']\bar{S}_{\g}(s',\mu')\right] ,
\label{g}
\\ \nonumber
 \hat{\o}(s,\mu)\equiv r_e{\o}(s,\mu)&=&-e^{(\s-\f{\g}{2})}\sum\limits_{n=1}^{\infty} 
\f{P^1_{2n-1}(\mu)}{2n(2n-1)\sin\th}
\left[\left(\f{1-s}{s}\right)^{2n+1}\int_0^s\f{s'^{2n}ds'}{(1-s')^{2n+2}}
\int_0^1d\mu'\sin\th'P^1_{2n-1}(\mu')\bar{S}_{\hat{\o}}(s',\mu') \right.
\\
&+&\left. \left(\f{s}{1-s}\right)^{2n-2}\int_s^1\f{(1-s')^{2n-3}ds'}{s'^{2n-1}}
\int_0^1d\mu'\sin\th'P^1_{2n-1}(\mu')\bar{S}_{\hat{\o}}(s',\mu')\right] .
\label{o}
\end{eqnarray}
where $P_n(\mu)$ are the Legendre polynomials and $P_n^m(\mu)$ are the 
associated Legendre polynomials and $\sin(n\th)$ is a function of $\mu$
through $\th=\cos^{-1}\mu$. The effective sources could be defined as
\begin{eqnarray}
 \bar{S}_{\s}(s,\mu)=r^2S_{\s}(s,\mu)
\\
 \bar{S}_{\g}(s,\mu)=r^2S_{\g}(s,\mu)
\\
 \bar{S}_{\hat{\o}}(s,\mu)=r_er^2S_{\o}(s,\mu)
\end{eqnarray}
The advantages of this Green's function approach for solving the 
elliptic field equations is that the asymptotic conditions on 
$\s, \g, \o$ are imposed automatically. The numerical integration 
of the Eqs. (\ref{r})-(\ref{o}) is straightforward. These
integrations are solved using the {\tt rns} code and we obtain the value of frame-dragging
precession rate inside the rotating neutron star using the equation 
(\ref{ltn}).

\subsection{Equation of state (EoS) of dense matter}
Recent observations of PSR J0348+0432 have reported the measurement of a 
2.01$\pm$0.04 M$_{\odot}$ neutron star \cite{anto}. This is the most accurately 
measured highest
neutron star mass so far. The accurately measured neutron star mass is a direct
probe of dense matter in its interior. This measured mass puts the strong 
constraint on the EoS.

Equations of state of dense matter are 
used as inputs in the calculation of frame-dragging in neutron star interior. 
We adopt three equations of state in this calculation. We are considering 
equations of state of $\beta$-equilibrated hadronic matter. The chiral EoS
is based on the QCD motivated chiral SU(3)$_L$ $\times$ SU(3)$_R$ model 
\cite{schm} and includes hyperons. We exploit the density dependent (DD) 
relativistic mean model to 
construct the DD2 EoS \cite{typ10}. Here the nucleon-nucleon interaction is 
mediated
by the exchange of mesons and the density dependent nucleon-meson
couplings are obtained by fitting properties of finite nuclei. The other EoS is
the Akmal, Pandharipande 
and Ravenhall (APR) EoS calculated in the variational chain summation method 
using Argonne $V_{18}$ nucleon-nucleon interaction and a fitted three nucleon 
interaction along with relativistic boost corrections \cite{apr}. 

We calculate the static mass limits of neutron stars using those three 
equations of state. Maximum masses and the corresponding radii of neutron 
stars are recorded in Table {\ref{table_0}}. Similarly maximum masses and 
the corresponding radii of rotating neutron stars at the mass shedding limits 
are also shown in the tables. These results show that maximum masses in all 
three cases are above 2 M$_{\odot}$ and compatible with the benchmark 
measurement mentioned above.

\section{Results and Discussion}

\begin{table}
\begin{center}
\begin{tabular}{|l|l|l|l|l|}
\hline
EoS&P(ms)&$\ve_c(10^{15}$ g/cm$^3$)&$M_G/M_{\odot}$& $R$ (km)\\
\hline\hline
APR& static & 2.78 & 2.190 & 9.93 \\
& 0.6291 & 1.50 & 2.397 & 14.53 \\
DD2& static & 1.94 & 2.417 & 11.90 \\
& 0.7836 & 1.00 & 2.677 & 17.53 \\
Chiral& static & 1.99 & 2.050 & 12.14 \\
& 0.8778 & 1.11 & 2.353 & 18.17 \\
\hline
\end{tabular}
\caption{Maximum gravitational masses $(M_G/M_{\odot})$, equatorial
radii $(R)$, and their corresponding central energy densities
$(\ve_c)$ for static $(\O=0)$ and Keplerian limit
$(P=P_K=2\pi/\O_K)$ with different EoS, where $P_K$ is the
Kepler period in millisecond.}
 \label{table_0}
\end{center}
\end{table}

We divide our results into two parts: in the first part
we show the frame-dragging effect in some pulsars which rotate with
fixed values of Kepler frequencies $\O_K$ and central densities $\ve_c$.
Next we consider pulsars whose masses and rotational periods are known from 
observations. Rotational frequencies of observed pulsars are
generally much lower than their Kepler frequencies $(\O<\O_K)$.

\subsection{Pulsars rotate with their Kepler frequencies $\O=\O_K$}

Figure {\ref{fig_apr} displays the frame-dragging frequency (or Lense-Thirring
precession frequency) as a function of
radial distance for APR EoS. Panel (a) of the figure represents the results 
along the equator whereas panel (b) implies those along the pole. For both 
panels of Fig. {\ref{fig_apr}}, we consider rotating neutron stars with 
central energy densities 5.2435 $\times 10^{14}$, 6.404 $\times 10^{14}$ and
7.534 $\times 10^{14}$ $g / cm^3$ and their corresponding Kepler frequencies
are 4000 (online-version: red), 5000 (online-version: green) and 6000 
(online-version: blue) $s^{-1}$, respectively whereas masses of the rotating 
compact stars in three cases range from $\sim$ 0.6 to $\sim$ 1.7 M$_{\odot}$. 
The Kepler periods $P_K=$1.57 ms, 1.26 ms and 1.05 ms correspond to the above
Kepler frequencies. Here and throughout 
the paper, $\O_{LT}$ is measured in a Copernican frame. For the
cases in panel (a), frame-dragging frequencies decrease initially with 
increasing distance from the centre and encounters a local minimum at a distance
$r_{min} \sim 0.5 r_e$ which is well below the surface. 
It is interesting to note here 
that the frame-dragging frequencies in all three cases rise again and attain 
a local maximum at the distance $r_{max} \sim 0.7 r_e$ and finally drop 
to smaller values at the surface. On the other hand, the frame-dragging 
frequencies along the pole smoothly vary from large values at the centre to 
smaller values at the surface as evident from panel (b) of Fig. {\ref{fig_apr}}.

Figure {\ref{fig_dd2}} shows the frame-dragging frequency along the equator
(panel (a)) and along the pole (panel (b)) for the DD2 EoS whereas 
Figure {\ref{fig_chi}} exhibits the frame-dragging frequency along the the
equatorial distance (panel (a)) and polar distance (panel (b)) for the Chiral 
EoS. In both figures, results are shown for Keplerian frequencies 
$\O_K =$ 4000, 5000 and 6000 $s^{-1}$.
However, the central energy densities corresponding to
the Keplerian frequencies mentioned above are different for three EoSs. 
The behaviour of frame-dragging frequencies along the equator and pole in
Fig. {\ref{fig_dd2}} and Fig. {\ref{fig_chi}} is qualitatively similar to the
results of Fig. {\ref{fig_apr}}. In Figs. {\ref{fig_apr}} - {\ref{fig_chi}},   
as the rotation frequency $\O_K$
and the central energy density $\ve_c$ increase, the frame-dragging frequencies 
increase and also the local maxima and minima
shift towards the surface of the neutron star
along the equator
for all three EoSs. It reveals an important conclusion that 
{\it the ratio of the positions 
of the local maxima and minima to the radius of the neutron star
must depend on $\O$ and $\ve_c$ for a particular pulsar.}

It could be easily seen from Eq. (\ref{ltn}) that the 
Lense-Thirring frequency inside a neutron
star is a function of both the radial distance $r$ and 
colatitude $\th$. The colatitude plays a major role to determine
the exact frame-dragging frequency at a particular point inside the 
rotating neutron star as evident from Figs. {\ref{fig_apr} - {\ref{fig_chi}. 
We obtain the Lense-Thirring frequency at the pole by just plugging-in $\th = 0$
in Eq. (\ref{ltn}) and it is given by $\O_{LT} = e^{-(\alpha+\sigma)}\omega$. 
It should be
noted here that the Lense-Thirring frequency is connected to $\omega$ which
appears as the non vanishing metric component in the metric of the rotating 
star.
According to the theorem by Hartle, the dragging of inertial frames as
represented by $\omega$ 
with respect to a distant observer decreases smoothly as a function of $r$ 
from a large value at the centre of the star to a smaller value at the surface
\citep{web} for both equatorial and polar cases. 
In this formalism the frame-dragging frequency depends solely on $r$.
For a fixed value of $r$, one gets the same frequency 
from the equator to the pole inside the rotating neutron star. 
We obtain the similar 
behaviour of the Lense-Thirring frequency along the pole in panel (b) of 
Figs. {\ref{fig_apr}} - {\ref{fig_chi}} as obtained in Hartle's formalism.} 
However, our results along the
equator are quite different from what was obtained using Hartle's formalism
\citep{web}. It is evident from Figs. \ref{fig_apr}-\ref{fig_chi} that the 
plots are smooth along the pole but not along the equator.
For the calculation of the Lense-Thirring frequency along the 
equator, we find that the second term of Eq. (\ref{ltn}) does not 
contribute. Further investigation of the first
term involving metric components $\sigma$, $\omega$
and their derivatives reveals that this term is responsible for the local
maxima and minima along the equator as reported above. 
The appearance of local maxima and minima in the 
Lense-Thirring frequencies along the equator may be attributed to the 
dependence of $\O_{LT}$ on $r$ and $\th$. As a consistency check,
we obtain two solutions for local maximum and minimum after
extremising the Eq. ({\ref{ltn}}) with respect to $r$. Details are given by
{Appendix \ref{apndx_2}}.

\begin{figure}
  \begin{center} 
\subfigure[along the equator]{
  \includegraphics[width=2in,angle=-90]{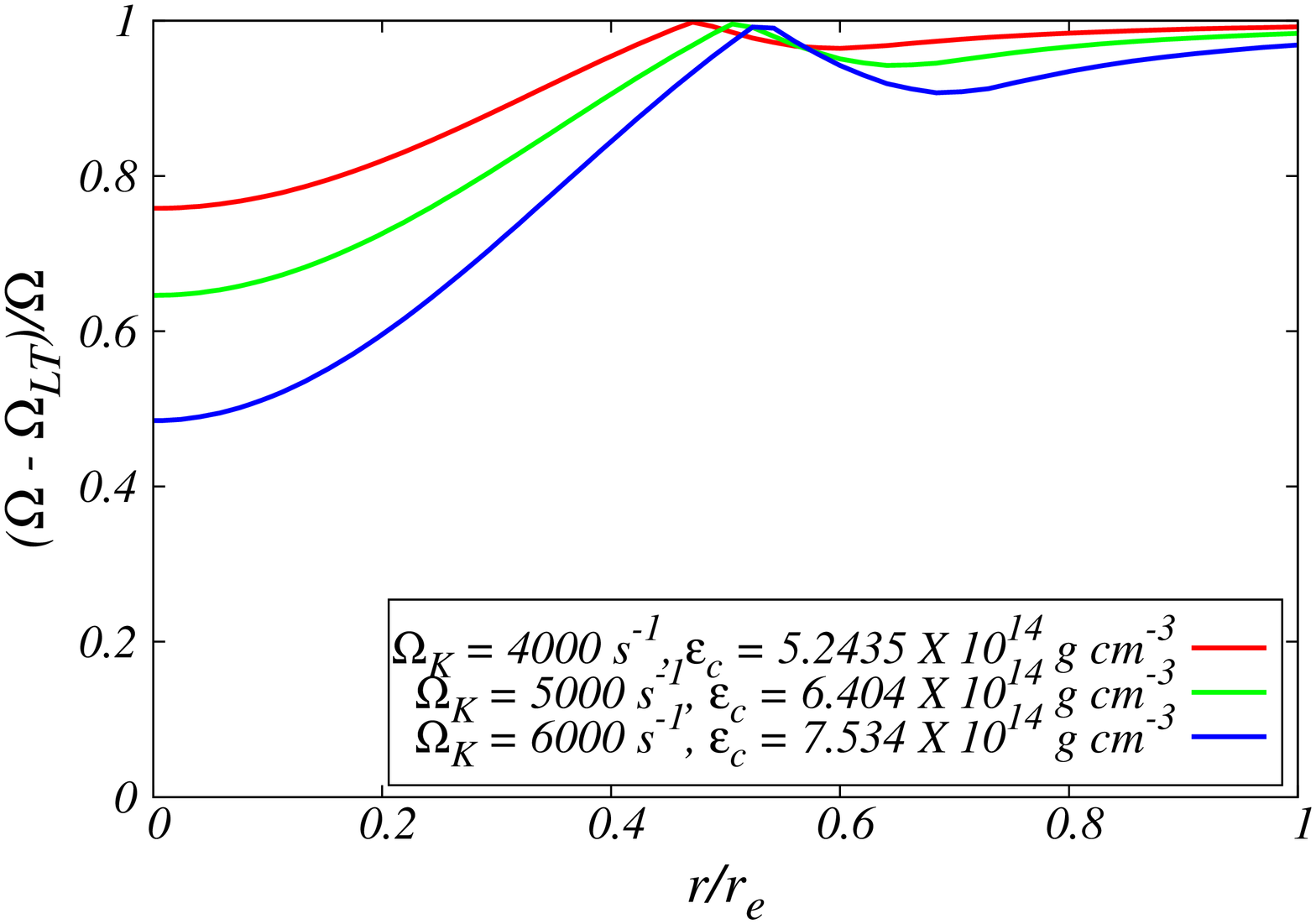}}
\subfigure[along the pole]{
\includegraphics[width=2in,angle=-90]{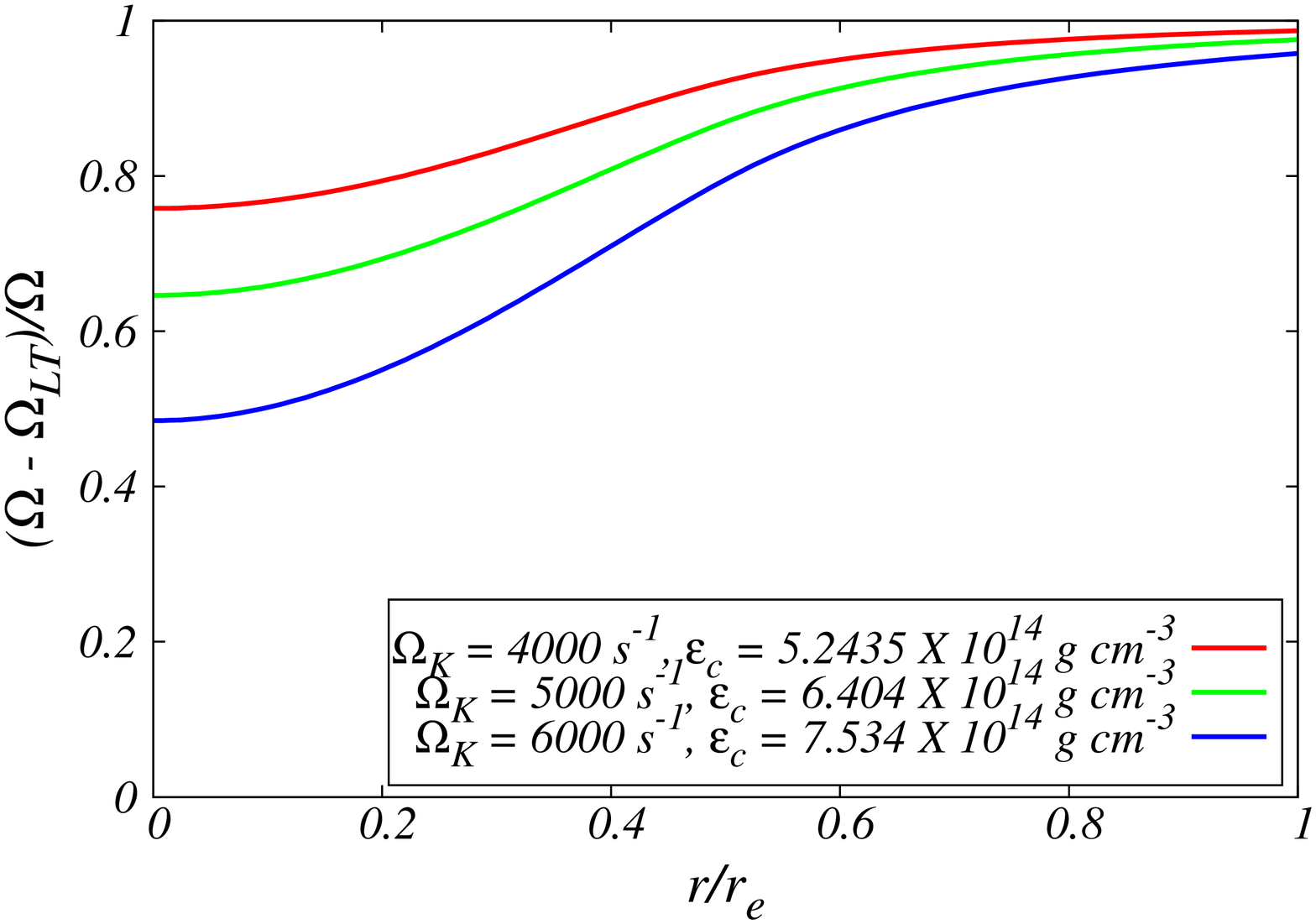}}
\caption{\label{fig_apr}\textit{Frame-dragging effect inside the rotating
      neutron stars from the origin to the surface,
calculated for the APR EoS. $\O_K$ and $\ve_c$ denote the Kepler 
frequency and the central star density, respectively. 
Surface of the neutron star along the pole
located around $0.6r_e$ but the plot is still valid beyond the surface 
of the pole as our formalism is applicable
for regions outside the pulsar.}}
\end{center}
\end{figure}

\begin{figure}
  \begin{center} 
\subfigure[along the equator]{
  \includegraphics[width=2in,angle=-90]{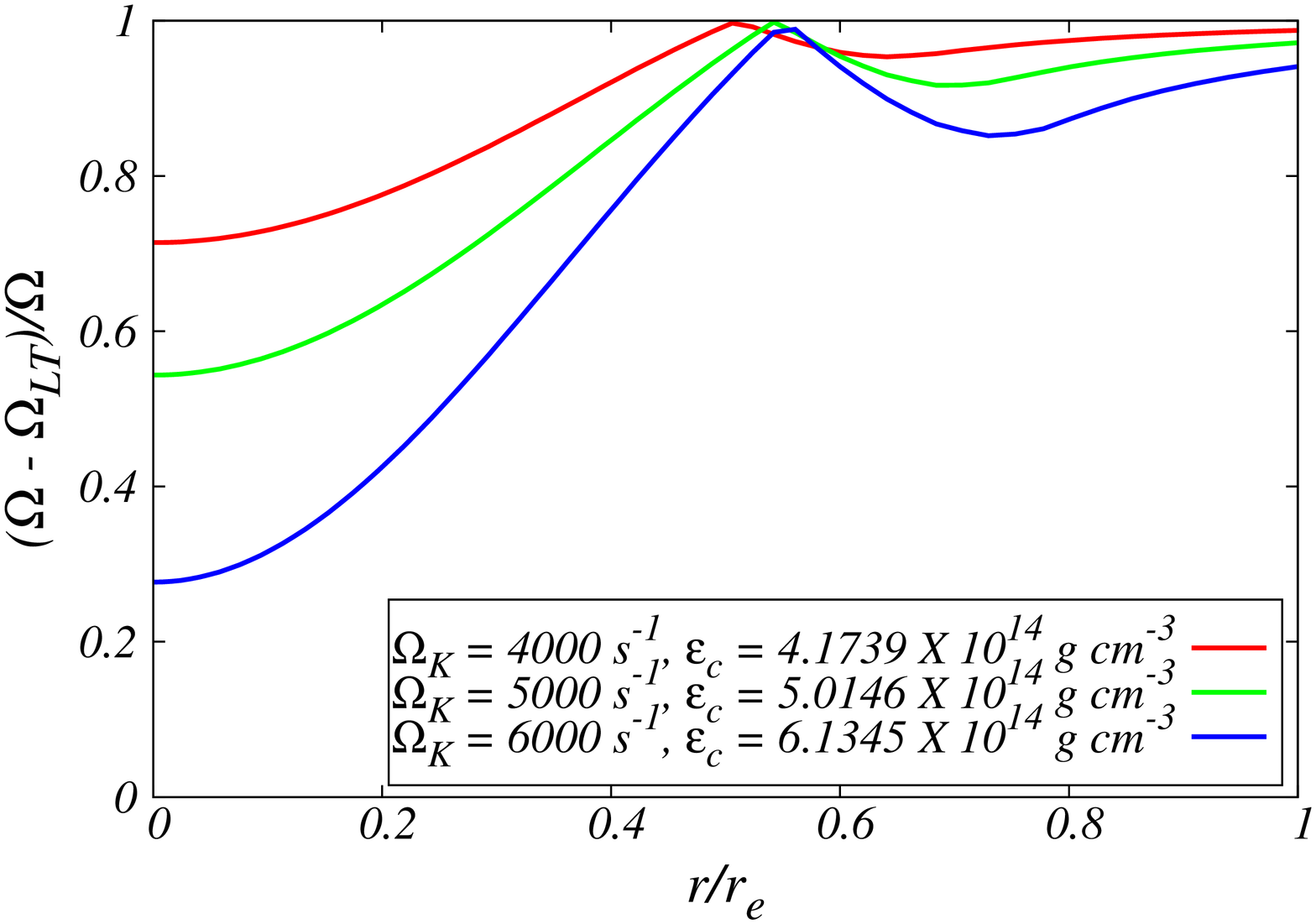}}
\subfigure[along the pole]{
\includegraphics[width=2in,angle=-90]{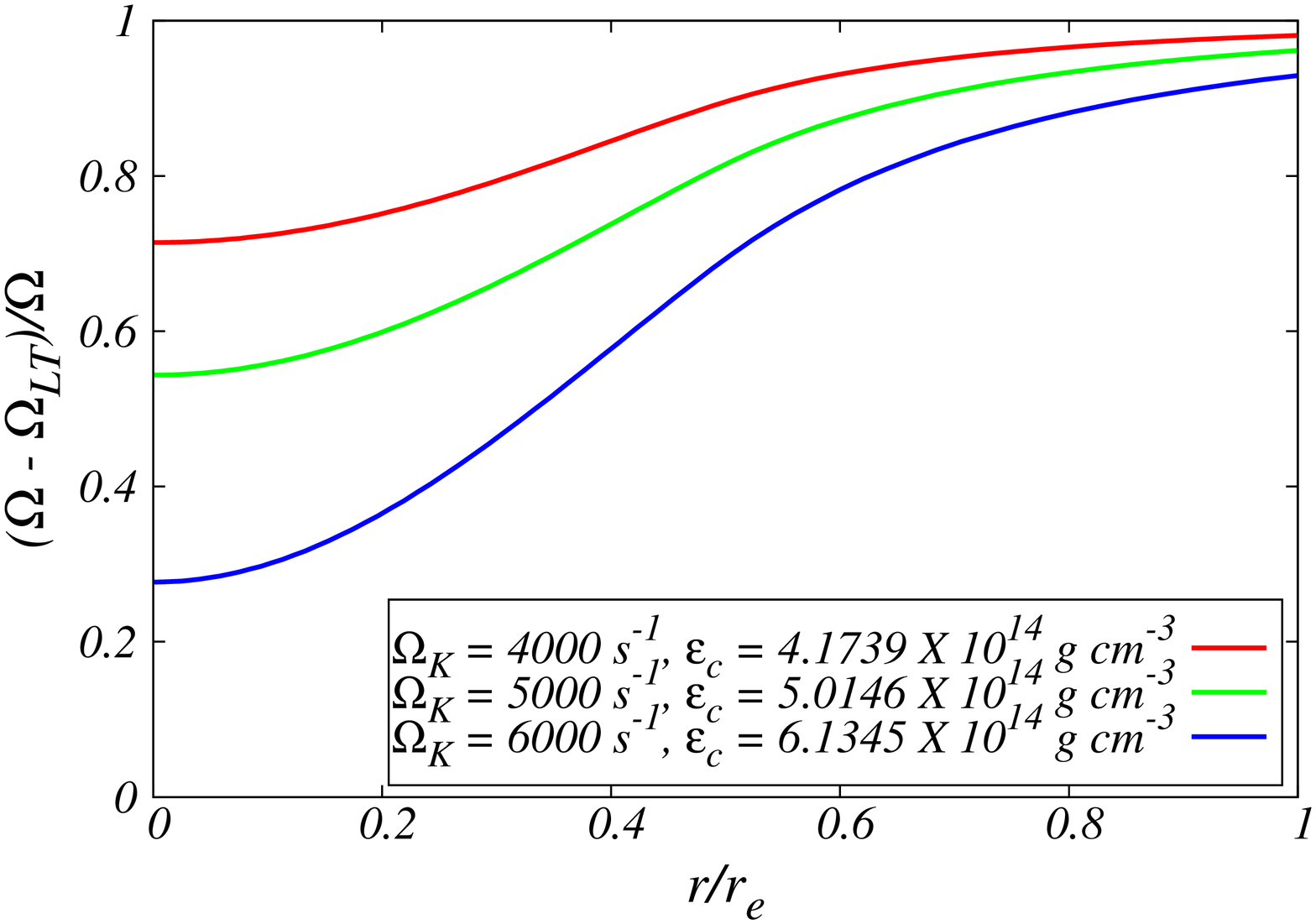}}
\caption{\label {fig_dd2}{\textit{Same as Fig. \ref{fig_apr}, but calculated for
the DD2 EoS}}}
\end{center}
\end{figure}

\begin{figure}
  \begin{center} 
\subfigure[along the equator]{
  \includegraphics[width=2in,angle=-90]{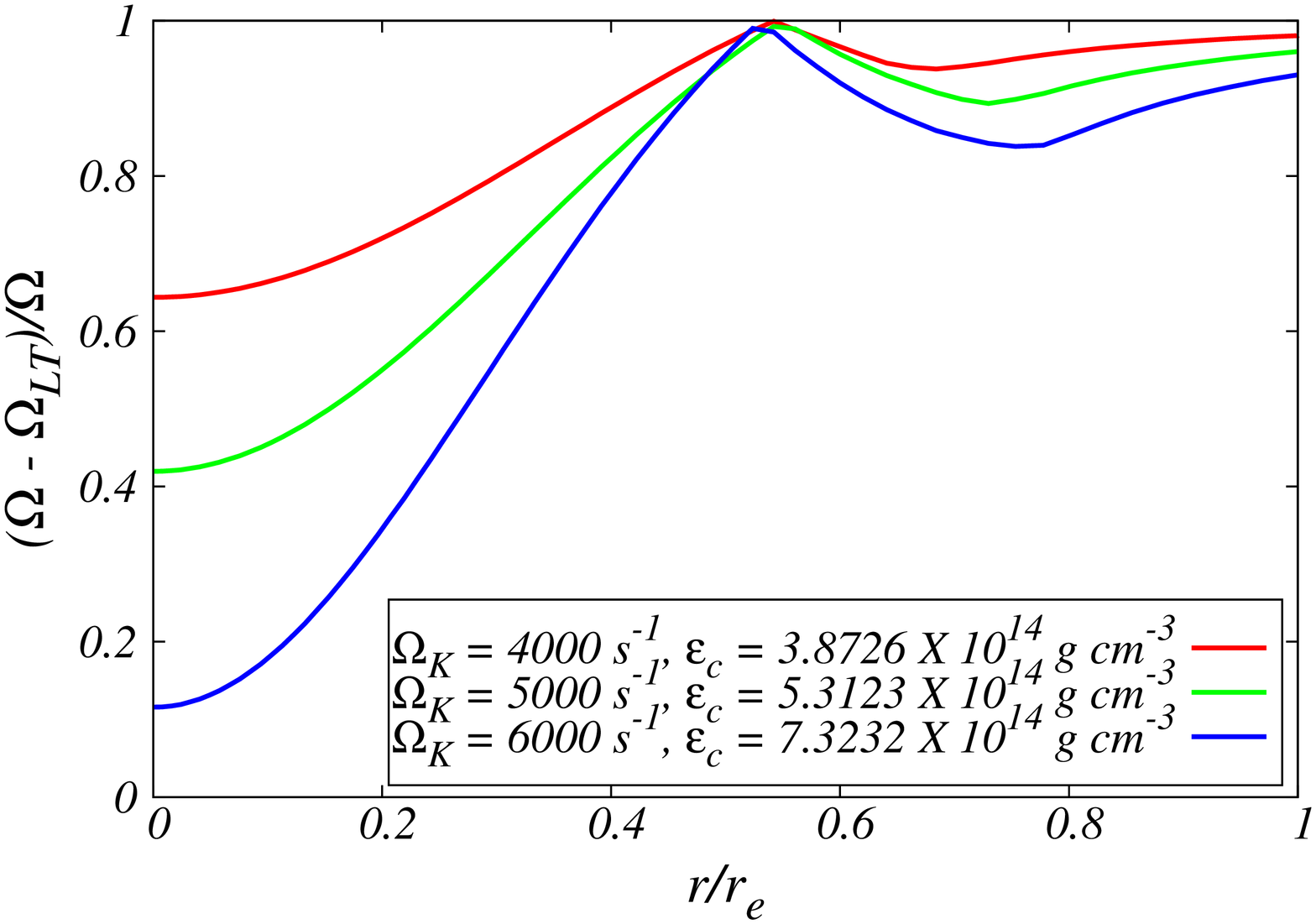}}
\subfigure[along the pole]{
\includegraphics[width=2in,angle=-90]{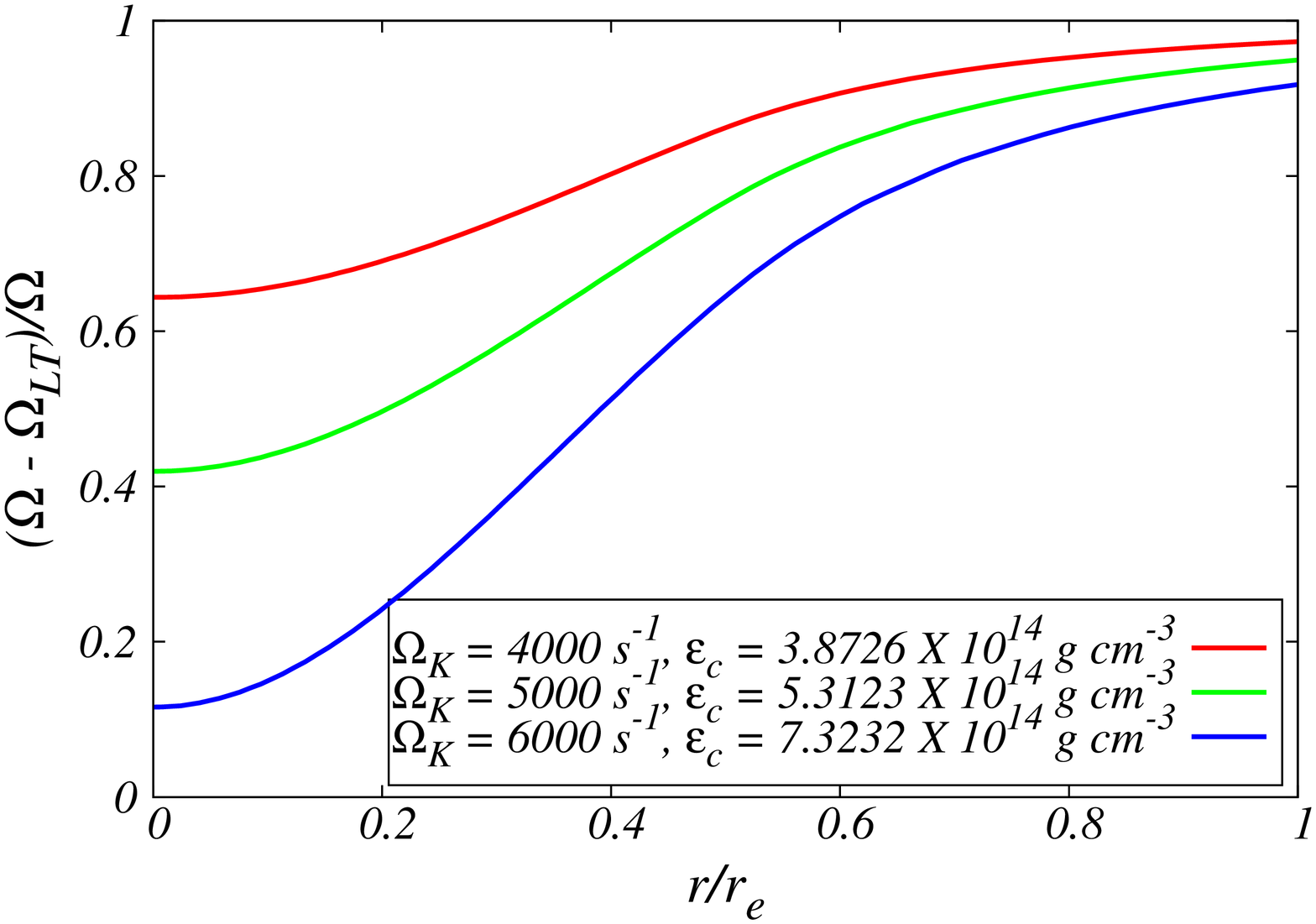}}
\caption{\label{fig_chi}{\textit{Same as Fig. \ref{fig_apr}, but calculated for
the Chiral EoS.}}}
\end{center}
\end{figure}

The normalised frame-dragging values at the centre 
$(\ti{\O}_c = {\O_{LT}^{centre}}/{\O})$ 
and the surface $(\ti{\O}_s = {\O_{LT}^{surface}}/{\O})$ 
of the star models with three EoSs are recorded in Table {\ref{table_1}}.
\begin{table}
\begin{center}
\begin{tabular}{|l|l||l|l|l||l|l|l|}
\hline
 &$P_K$&\multicolumn{3}{l|}{Along the equator}
&\multicolumn{3}{l|}{Along the pole}\\
\cline{3-8}
&(ms)&APR&DD2&Chiral&APR&DD2&Chiral\\
\hline\hline
& 1.57 & 0.008 & 0.013 & 0.019 & 0.046 & 0.069 & 0.099 \\
$\ti{\O}_s$& 1.26 & 0.016 & 0.029 & 0.040 & 0.087 & 0.139 & 0.184 \\
& 1.05 & 0.031 & 0.059 & 0.070 & 0.151 & 0.252 & 0.287 \\
\hline
& 1.57 & 0.242 & 0.286 & 0.356 & 0.242 & 0.286 & 0.356 \\
$\ti{\O}_c$ & 1.26 & 0.354 & 0.457 & 0.580 & 0.354 & 0.457 & 0.580 \\
& 1.05 & 0.515 & 0.723 & 0.884 & 0.515 & 0.723 & 0.884 \\
\hline
\end{tabular}
\caption{Normalised angular velocities of the local inertial frame-dragging
at the surface $\ti{\O}_s$ and the centre 
$\ti{\O}_c$  of the neutron stars which are rotating 
at their respective Kepler periods $(P_K\equiv 2\pi/\O_K)$ 
as measured by a distant observer.}
 \label{table_1}
\end{center}
\end{table}
It is noted that the normalised frame-dragging value at the star's center 
is maximum and 
falls off on the surfaces of the equator and pole 
for three EoSs irrespective of whether the compact star is rotating slowly or
fast. However, for a particular EoS, the normalised frame-dragging value 
at the star's centre and surface is higher 
for a fast rotating star with $P_K = 1.05$ ms than 
those of a slowly rotating star with $P_K = 1.57$ ms for both cases along the 
equator and pole. One can see another interesting 
thing from the Table \ref{table_1} that $\ti{\O}_s$ is always higher 
at the pole than $\ti{\O}_s$ at the equator for a particular 
pulsar. It is due to the effect of rotation frequency $\O$
(of the star) for which pole is nearer to the center than
the surface as it is evident from Table \ref{table_1}.
 Thus, the inertial frame-dragging effect is
higher at the
surface of the neutron star along the pole than that 
at the surface of the neutron star along the equator.

Now we investigate the dependence of local maxima 
and minima in $\O_{LT}$ along
the equator on the angle $\th$. 
In Figure {\ref{cr}}, $\O_{LT}$ is shown as a
function of $s$ defined by Eq. (\ref{rs}) and $\cos\th$ for DD2 (panel(a)) and Chiral
(panel (b)) EoSs.  
The local maxima and minima in $\O_{LT}$ along the equator are clearly visible
in both panels of Fig. {\ref{cr}}. 
It has been already noted that $\O_{LT}$ 
along the pole decreases smoothly from the center to the surface. As 
the latitude ($\th'=\pi/2-\th$) increases
from the equator to the pole, the height between the maximum and minimum of
$\O_{LT}$ diminishes and after a certain `critical' angle ($\th_{cr}$) both 
extrema disappear and the plot is smooth like the plot along
the pole. The critical angle could be seen from the 3-D plot in Fig. \ref{cr}.
\begin{figure}
\begin{center} 
\subfigure[$DD$2 EoS]{
\includegraphics[width=2in,angle=-90]{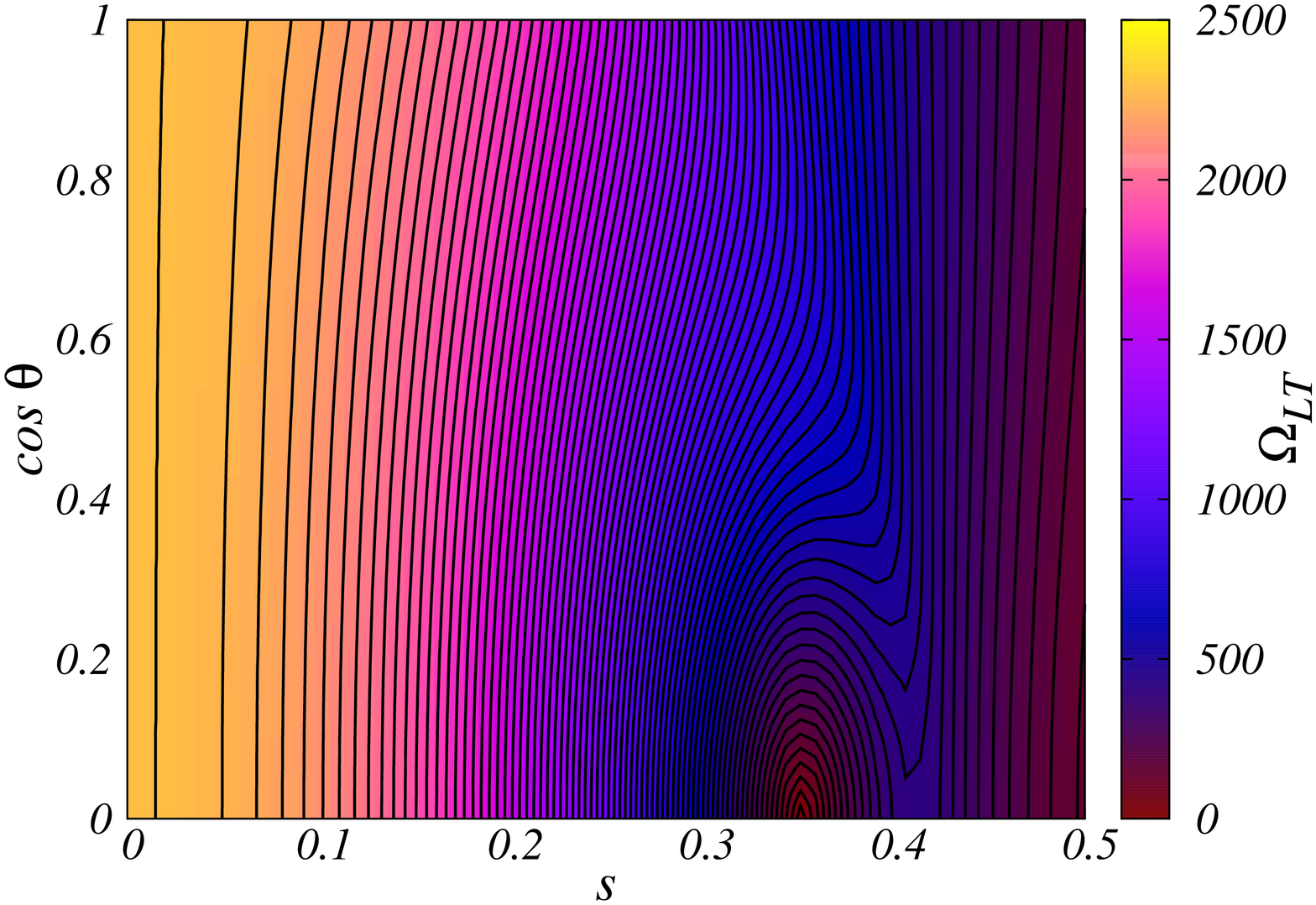}}
\subfigure[APR EoS]{
\includegraphics[width=2in,angle=-90]{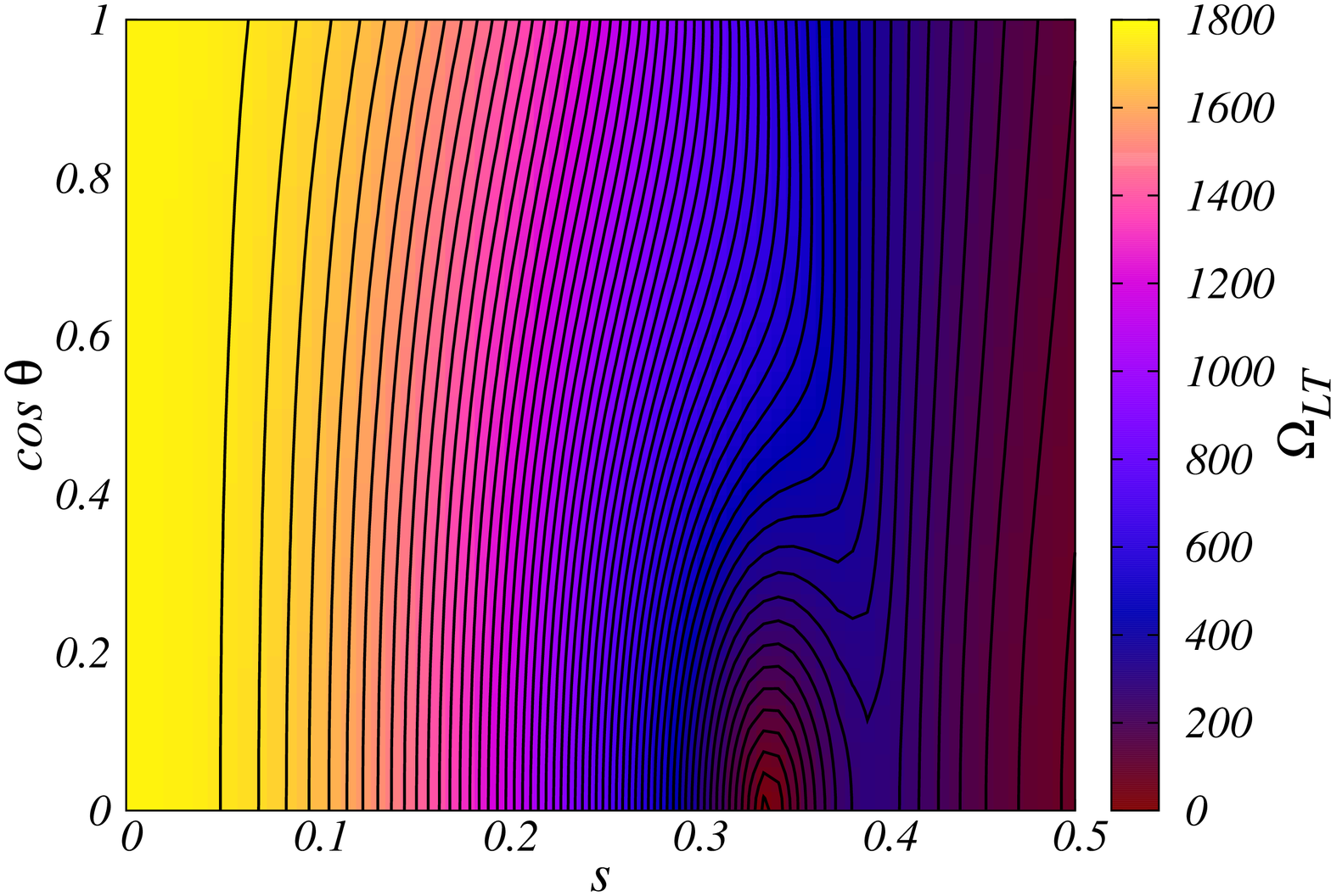}}
\caption{\label{cr}{\textit{3-D plots of $\O_{LT}$ of the pulsar which is
rotating with $\O_K=5000\, \text{s}^{-1}$} as a function of $s$ and $\cos\th$
for (a) DD2 EoS and (b) APR EoS.}}
\end{center}
\end{figure}
This value of the critical angle is $\mu\approx 0.5$ or $\th_{cr}'=30^{\circ}$ 
where local maximum  $(r_{max})$ and minimum $(r_{min})$ disappear. So, if we plot 
$\O_{LT}$ vs $\th$ at the point $r_{max}$ for a specific Kepler frequency, 
(namely 
$\O_K=5000\, \text{s}^{-1}$), we could find that the frame-dragging frequency 
increases from the equator to the pole for the specific $r_{max}$ as 
it is exhibited by Fig. {\ref{cr}}.

\subsection{Pulsars rotate with their frequencies $\O<\O_K$}

Now we apply our exact formula of $\O_{LT}$ to three known pulsars.
Three pulsars chosen for this purpose are J1807-2500B, J0737-3039A and
B1257+12. Periods of those pulsars are given by Table {\ref{table_2}}.  
Masses of those pulsars are also known and range from 1.337 to 1.5 M$_{\odot}$.
Furthermore, we adopt the same EoSs in this calculation as considered in
the previous sub-section. Though periods of these pulsars are larger than 
the Keplerian periods, the calculation of $\O_{LT}$ inside these real
pulsars are equally important like the cases with Kepler frequencies 
demonstrated already. 

We calculate the normalised angular velocity at the centre and surface of these 
pulsars with APR, DD2 and Chiral EoSs. It is noted from Table \ref{table_2}
that the behaviour of the normalised angular velocity from the centre to the
surface or along the pole and equator is qualitatively same as shown in 
Table {\ref{table_1}}.  
\begin{table}
\begin{center}
\begin{tabular}{|l|l|l||l|l|l||l|l|l|}
\hline
&Name of&$P$&\multicolumn{3}{l|}{Along the equator}
&\multicolumn{3}{l|}{Along the pole}\\
\cline{4-9}
&the Pulsar&(ms)&APR&DD2&Chiral&APR&DD2&Chiral\\
\hline\hline
& J1807-2500B & 4.19 & 0.099 & 0.075 & 0.064 & 0.156 & 0.120 & 0.105 \\
$\ti{\O}_s$& J0737-3039A & 22.70 & 0.095 & 0.073 & 0.062 & 0.154 & 0.122 & 0.106 \\
& B1257+12 & 6.22 & 0.122 & 0.091 & 0.077 & 0.188 & 0.145 & 0.126 \\
\hline
& J1807-2500B & 4.19 & 0.707 & 0.548 & 0.516 & 0.707 & 0.548 & 0.516 \\
$\ti{\O}_c$ & J0737-3039A & 22.70& 0.685 & 0.538 & 0.502 & 0.685 & 0.538 & 0.502 \\
& B1257+12 & 6.22 & 0.825 & 0.632 & 0.601 & 0.825 & 0.632 & 0.601 \\
\hline
\end{tabular}
\caption{Normalised angular velocities of the local inertial 
frame-dragging at the surface $\ti{\O}_s$ and the centre
$\ti{\O}_c$ of some known rotating neutron stars.}
 \label{table_2}
\end{center}
\end{table}

We also plot the frame-dragging frequency ($\O_{LT}$) as a function of radial 
distance
along the equator (panel (a)) and pole (panel (b)) for three
EoSs in Figures {\ref{j1807}} - {\ref{b1257}}. 
The frame-dragging frequency behaves smoothly along the pole from the centre to 
the surface as shown by panel (b) of these figures.
Results of panel (a) of the figures show
similar features of local maxima and minima along the equator as found in 
Figs. \ref{fig_apr}-\ref{fig_chi}.  
We note that all the local minima of $\O_{LT}$ are located around 
$r_{min}\sim 0.7r_e$
and the local maxima are located around $r_{max}\sim 0.9r_e$
in Figs. {\ref{j1807}} - {\ref{b1257}}.
\begin{figure}[h!]
\begin{center}
\subfigure[along the equator]{
\includegraphics[width=2in,angle=-90]{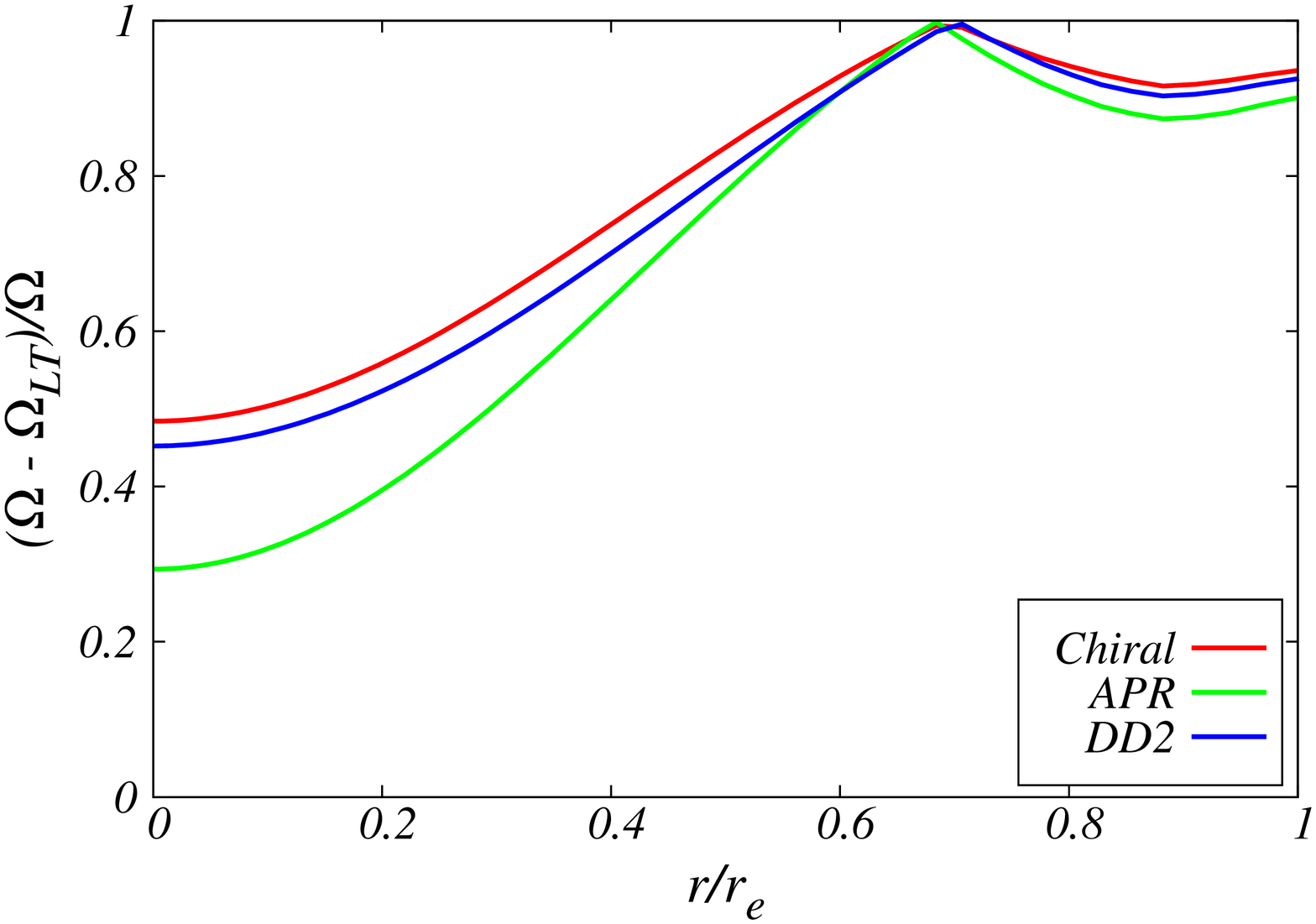}
}
\subfigure[along the pole]{
\includegraphics[width=2in,angle=-90]{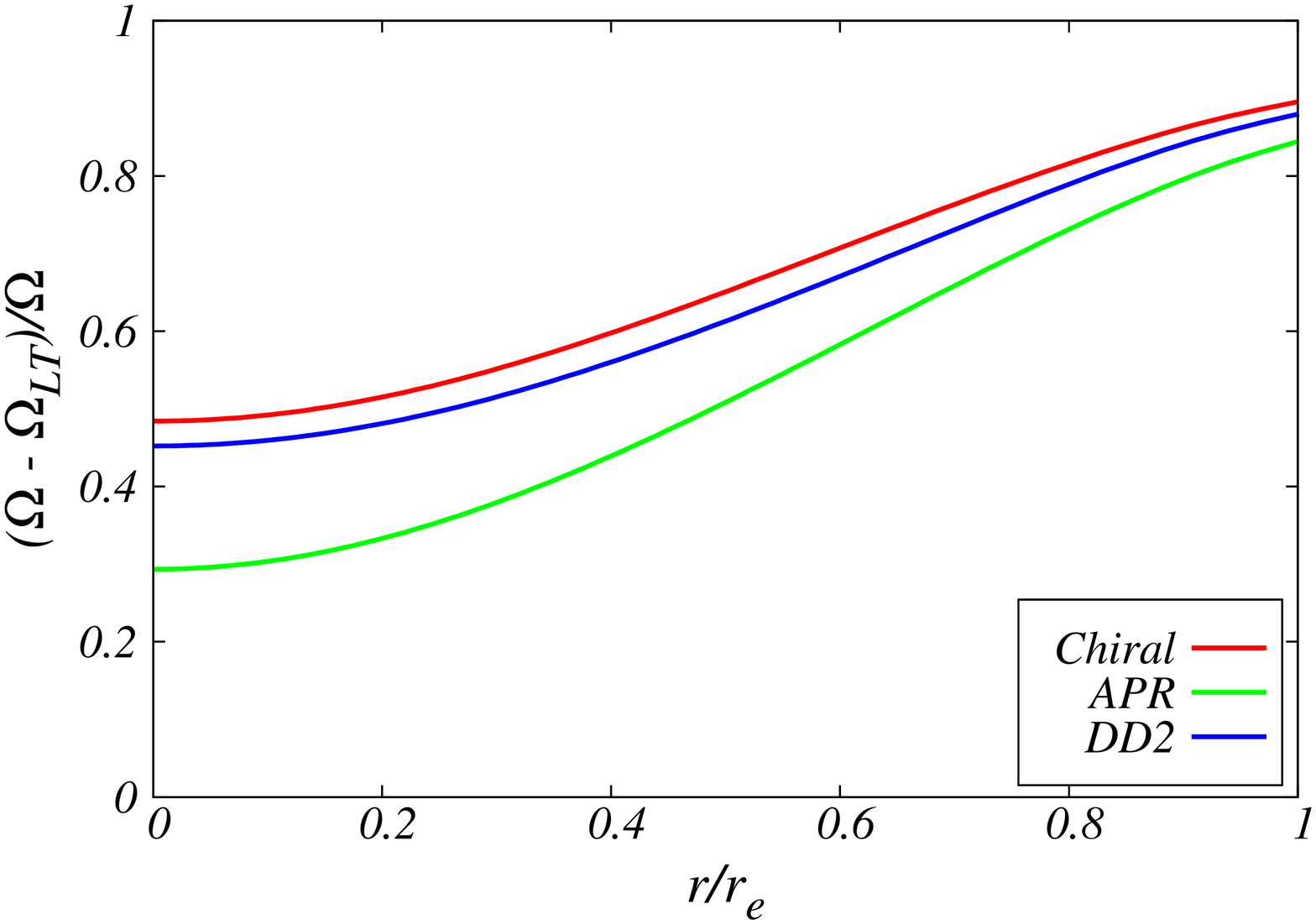}
}
\caption{\label{j1807} \textit{Frame-dragging effect inside the rotating
      neutron star from the origin to the surface,
calculated for J1807-2500B ($M=1.366 M_{\odot}, \O = 1500.935\, \text{s}^{-1}$)}}
\end{center}
\end{figure}

\begin{figure}[h!]
\begin{center}
\subfigure[along the equator]{
\includegraphics[width=2in,angle=-90]{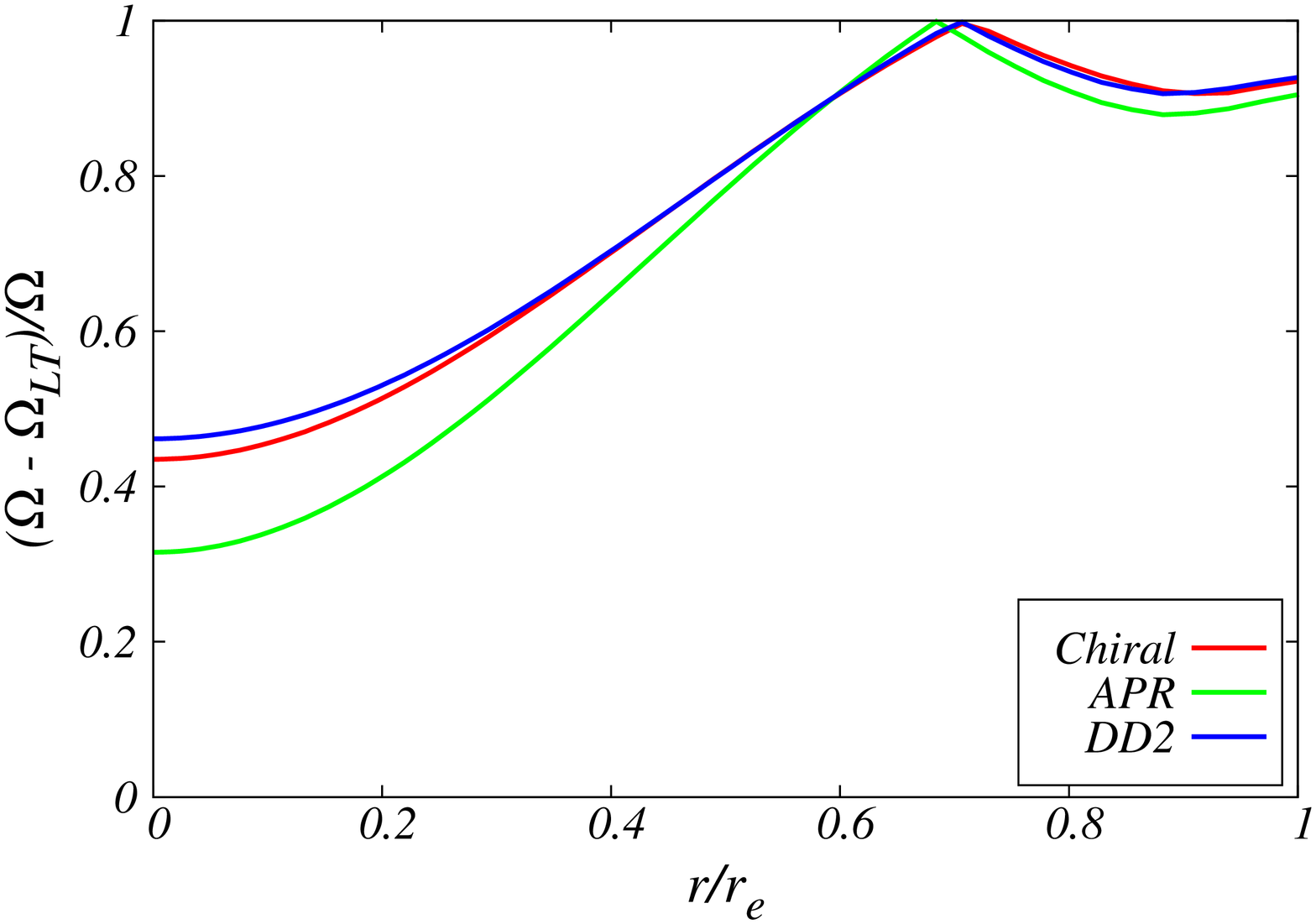}
}
\subfigure[along the pole]{
\includegraphics[width=2in,angle=-90]{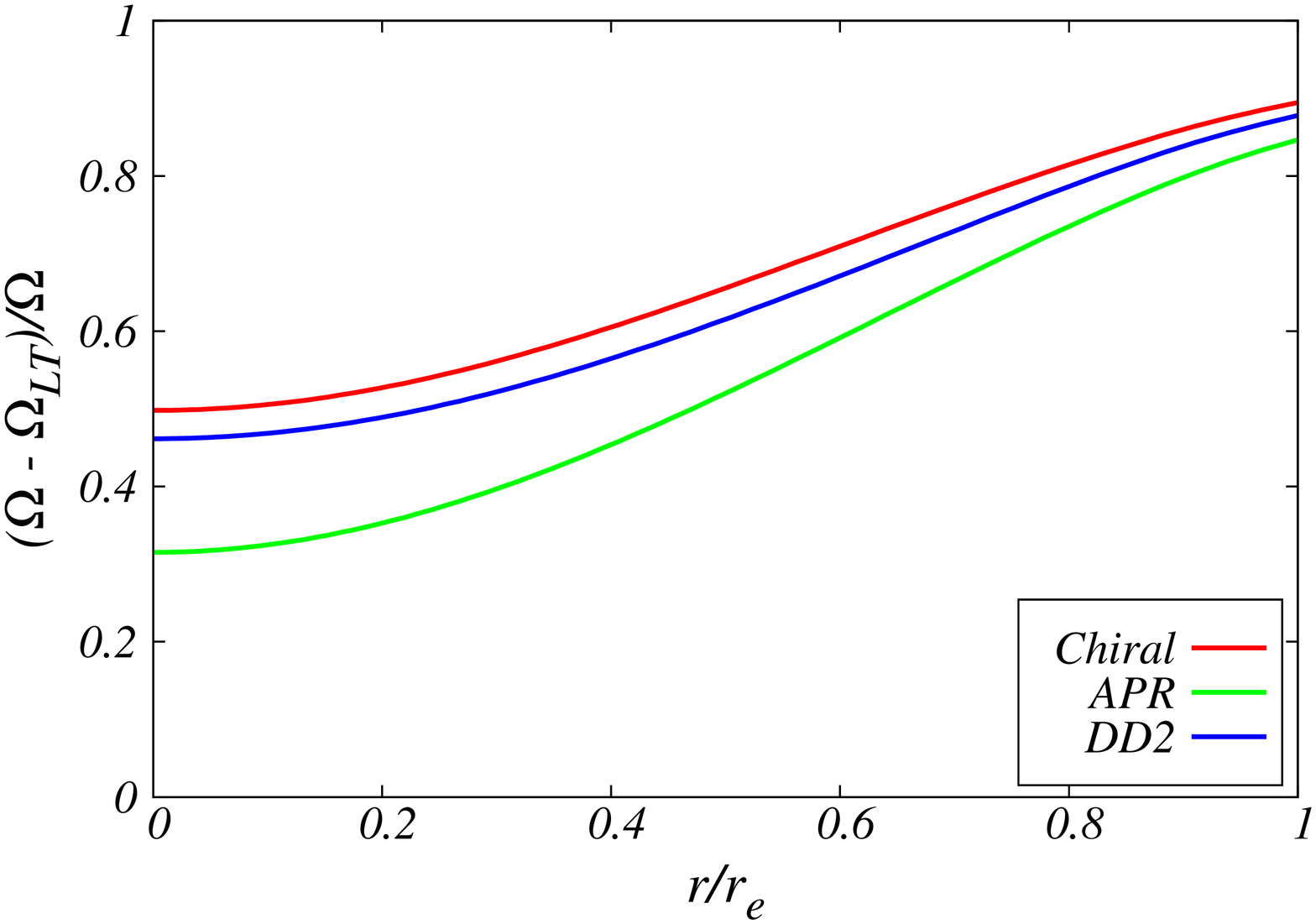}
}
\caption{\label{j0737} \textit{Frame-dragging effect inside the rotating
      neutron star from the origin to the surface,
calculated for J0737-3039A ($M=1.337 M_{\odot}, \O = 276.8\, \text{s}^{-1}$)}}
\end{center}
\end{figure}

\begin{figure}[h!]
\begin{center}
\subfigure[along the equator]{
\includegraphics[width=2in,angle=-90]{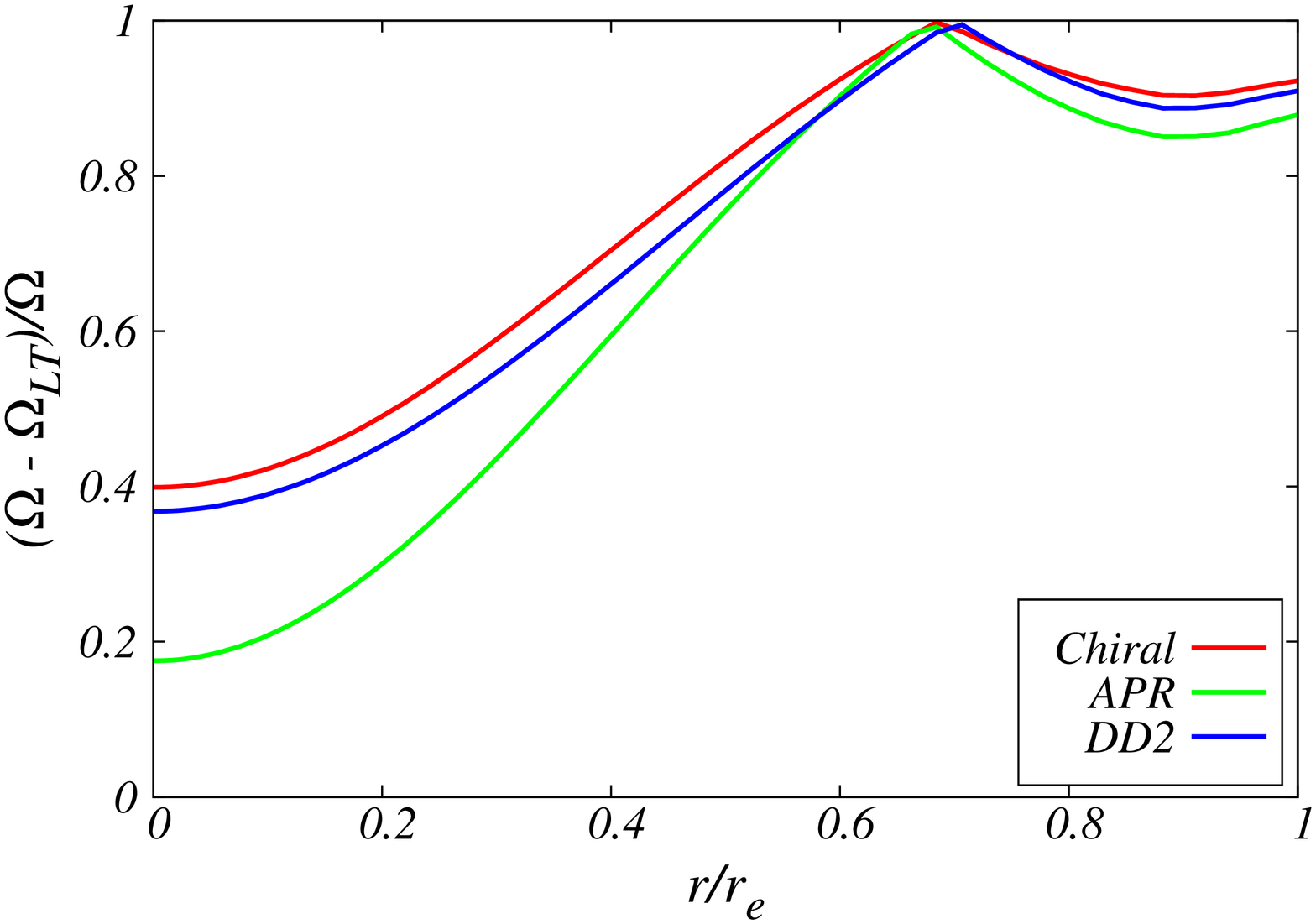}
}
\subfigure[along the pole]{
\includegraphics[width=2in,angle=-90]{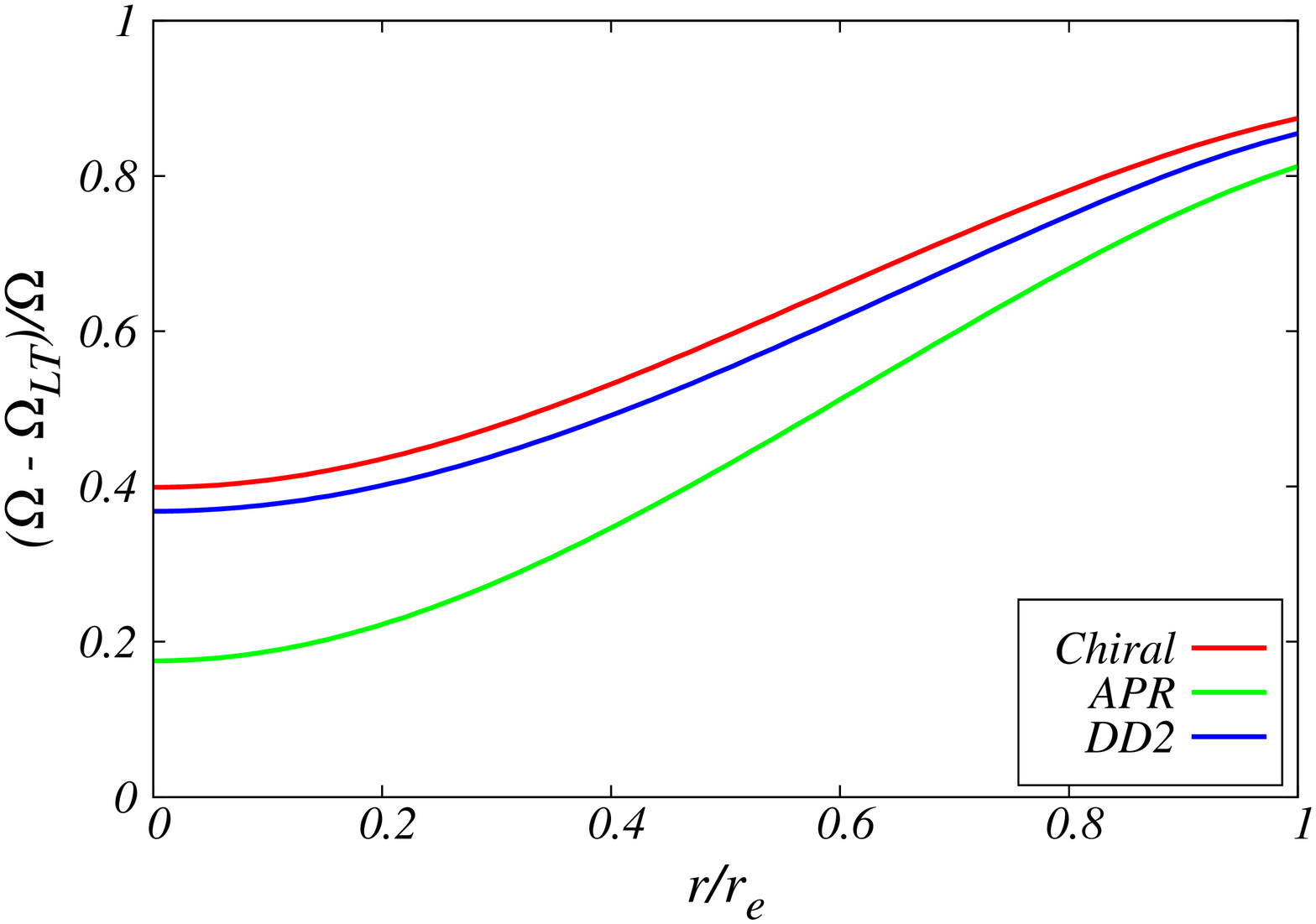}
}
\caption{\label{b1257} \textit{Frame-dragging effect inside the rotating
      neutron star from the origin to the surface,
calculated for B1257+12 ($M=1.5 M_{\odot}, \O = 1010.321\, \text{s}^{-1}$)}}
\end{center}
\end{figure}
We also plot the $\O_{LT}$ of pulsar J0737-3039A as a function of $s$ and 
$\cos\th$ for DD2 (panel (a)) and APR (panel (b)) EoSs in Figure 
{\ref{cr2}. It is noted 
from Fig. \ref{cr2} that the value of 
$\th_{cr}$ is around $30^{\circ}$ for the pulsar J0737-3039A
for DD2 and APR EoSs. 

\begin{figure}
  \begin{center}
\subfigure[$DD$2 EoS]{
\includegraphics[width=2in,angle=-90]{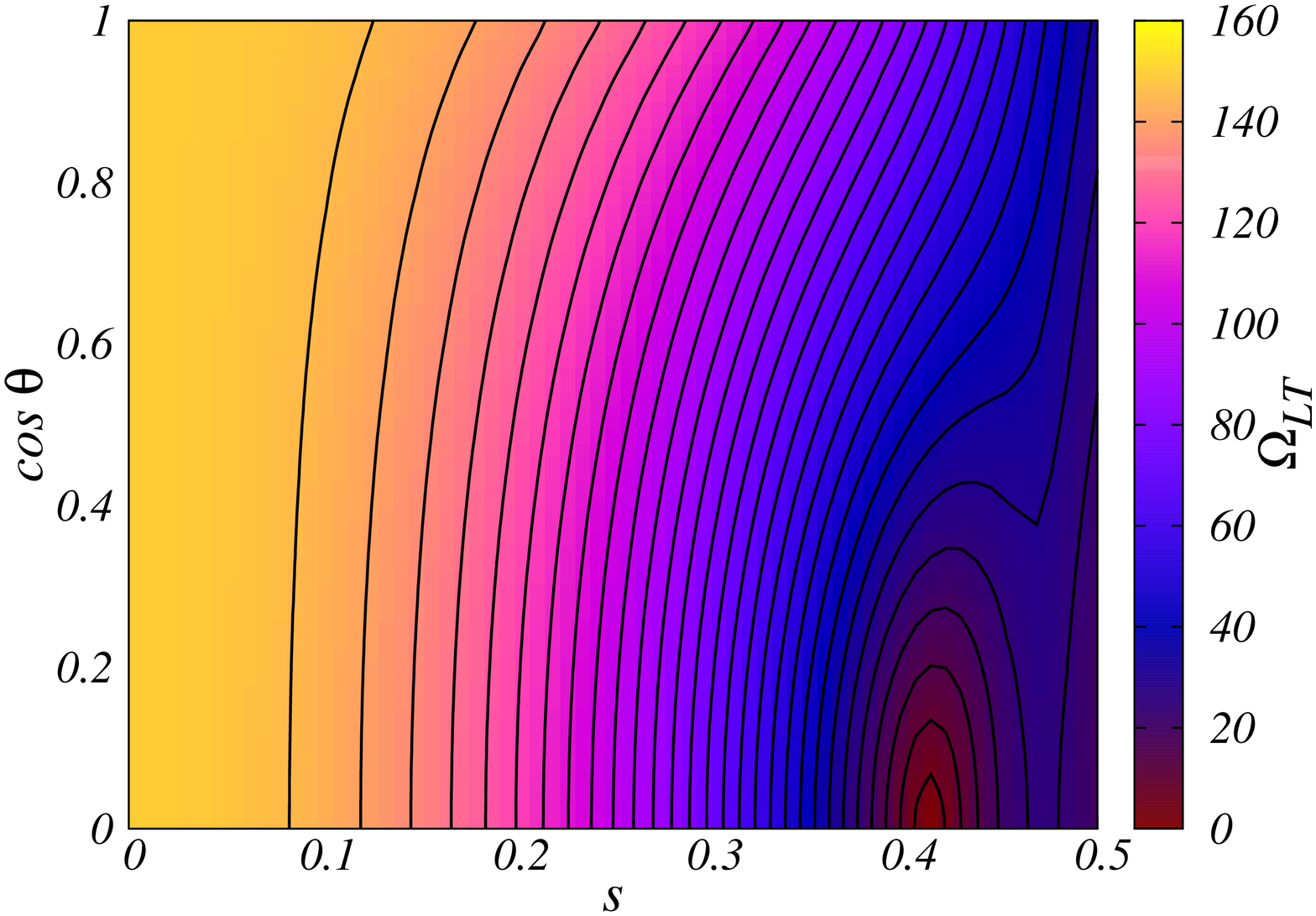}} 
\subfigure[APR EoS]{
 \includegraphics[width=2in,angle=-90]{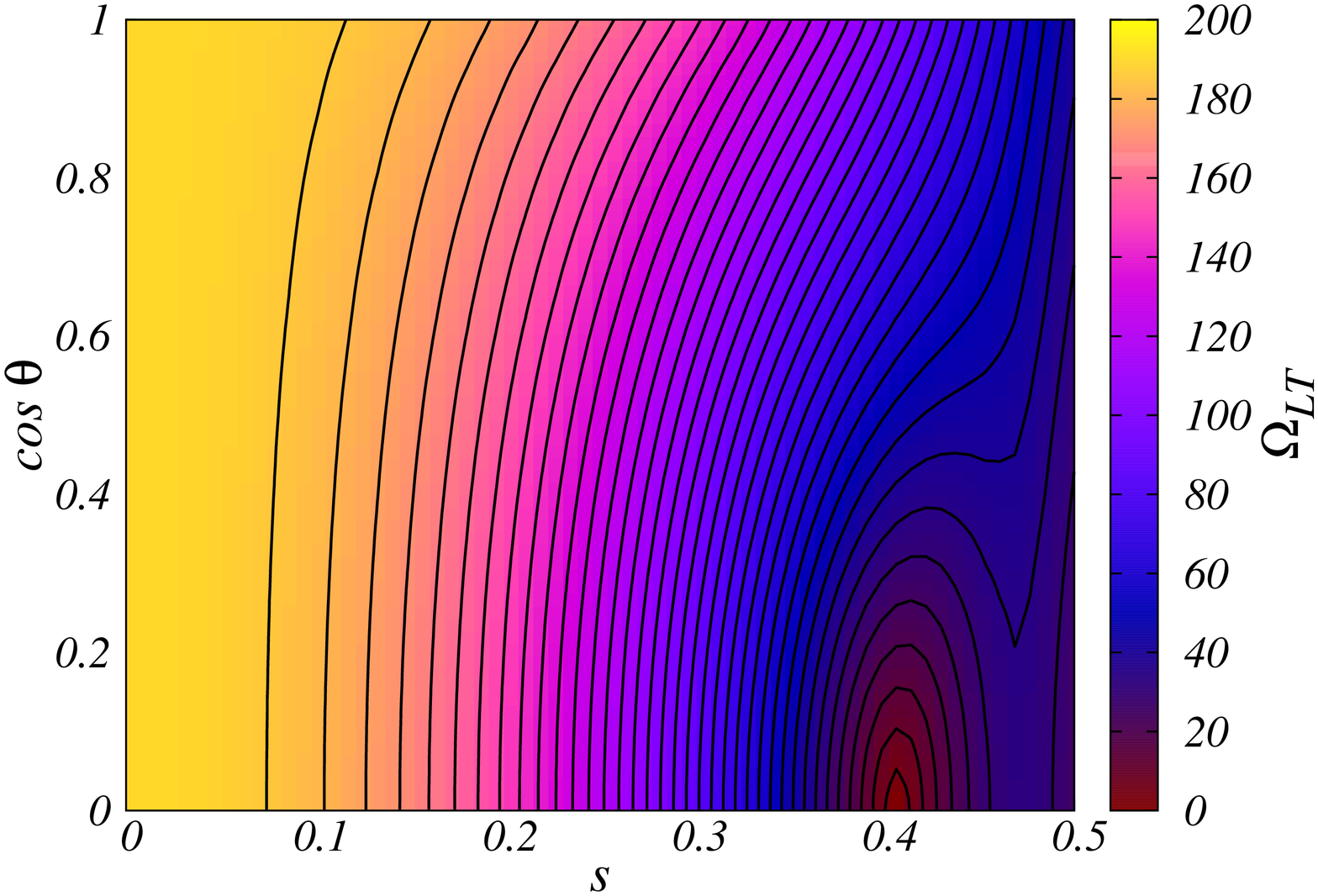}}
\caption{\label{cr2}\textit{3-D plots of $\O_{LT}$ of the pulsar J0737-3039A
as a function of $s$ and $\cos\th$ for DD2 (panel (a)) and APR (panel (b))
EoSs.}}
\end{center}
\end{figure}

\section{Summary}
We have derived the exact frame dragging 
frequency inside the rotating neutron star without making
any assumption on the metric components
and energy-momentum tensor.
We show that the frequency must depend both on $r$ and $\th$.
It may be recalled that the frame-dragging frequency depends only on $r$ in
Hartle's formalism because of the slow rotation approximation. We predict the 
exact frame-dragging
frequencies for some known pulsars as well as neutron stars rotating at their
Keplerian frequencies. We have also estimated
Lense-Thirring precession frequencies at the centers
of these pulsars without imposing any boundary conditions 
on them. We have found local maxima and minima along the equator 
due to the dependence of $\O_{LT}$ on the colatitude ($\th$) 
inside pulsars. 
The positions of local maximum and minimum depend on the frequency
$\O$ and the central 
density $\ve_c$ of the particular pulsar. Furthermore, it is observed that 
local maximum and minimum in $\O_{LT}$ along the equator disappear at a 
critical angle $\th_{cr}$.

Quasi periodic oscillations (QPOs) in magnetars were studied by various 
groups. 
These studies in several cases were carried by considering spherical and 
non-rotating relativistic stars having dipolar magnetic field configuration
\citep{sota07}. It would be worth investigating this problem for rotating
relativistic stars. In particular, we are studying the effect of our exact
frame-dragging formulation on the magnetic field distribution in the star and
its implications on QPOs. This will be published in future.  

{\bf Acknowledgements:} CC and KPM would like to thank 
Prof. Dr. P. Majumdar for various discussions 
regarding this project. 
His comments and valuable suggestions help CC a lot
to make this work more appropriate. CC and KPM also thank
to Mr. P. Char and Mr. A. Kheto for important discussions
regarding this project.
Last but not the least, CC thanks Prof. Dr. K.D. Kokkotas of 
University of T\"ubingen, Germany for gracious hospitality 
during an academic visit and for invaluable discussions 
regarding the subject of this paper. Two of us (KPM \& CC) are grateful to
Dept. of Atomic Energy (DAE, Govt. of India) for financial assistance.

\begin{appendix}
\section{Appendix: Consistency check of local maximum and minimum in $\O_{LT}$}

To find out the local maximum and minimum in $\O_{LT}$
we differentiate Eq.(\ref{ltn}) 
with respect to $r$ and obtain
\begin{eqnarray}\nonumber
\f{d{\O}_{LT}}{dr}&=&\O_{LT}\left[-(\a_{,r}+\s_{,r})-2\f{r\o\sin^2\th(r\o_{,r}+\o)-\s_{,r}e^{2\s}}
{\o^2r^2\sin^2\th-e^{2\s}}\right]+\f{1}{\O_{LT}}.\f{e^{-2(\a+\s)}}{4(\o^2r^2\sin^2\th-e^{2\s})^2}
\\ \nonumber
 &.&\left\{A\sin^2\th \left[\o r^2\sin^2\th(3\o\o_{,r}+2r\o_{,r}^2+r\o \o_{,rr})
 +2\s_{,r}e^{2\s}(2\o+r\o_{,r}-2\o r\s_{,r})\right.\right.
\\ 
&+& \left.\left.e^{2\s}(3\o_{,r}+r\o_{,rr}-2r\o_{,r}\s_{,r}-2\o\s_{,r}-2r\o\s_{,rr})\right]
+B \left[r\o\sin^3\th(2\o\o_{,\th}+2r\o_{,r}\o_{,\th}+r\o\o_{,\th r})\right.\right.
\\ \nonumber
&+&\left.\left.2\s_{,r}e^{2\s}(2\o\cos\th+\o_{,\th}\sin\th-2\o\s_{,\th}\sin\th)
+e^{2\s}(2\o_{,r}\cos\th+\o_{,\th r}\sin\th-2\sin\th(\o\s_{,\th r}+\o_{,r}\s_{,\th}))\right]\right\}
\label{diff_ltn}
\end{eqnarray}
in where
\begin{eqnarray}
A&=&r^3\o^2\o_{,r}\sin^2\th+e^{2\s}(2\o+r\o,_r-2\o r \s,_r) \,\,\, ,
\\
B&=&r^2\o^2\o_{,\th}\sin^3\th+e^{2\s}(2\o \cos\th+\o_{,\th}\sin\th
-2\o\s_{,\th}\sin\th)
\\
\text{and}&& \nonumber
\\
&&\o_{,\th r}\equiv \f{\p^2\o}{\p\th\p r}\nonumber
\end{eqnarray}
 Setting $\f{d\O_{LT}}{dr}|_{(r=R_0, \th=\pi/2)}=0$ and solving it numerically
in the region $0<R_0<r_e$,  
we obtain two positive real roots of $r$ inside the rotating
neutron star. One of these is $R_{01}=r_{max}$ and another is $R_{02}=r_{min}$.
These are basically the {\it local maximum} 
and {\it local minimum} of the function $\O_{LT}$ along the equator.
These local maximum and local minimum are absent for 
$\f{d\O_{LT}}{dr}|_{(r=R_0, \th=0)}=0$. Thus, we cannot see
any such extremum for the plots of $\O_{LT}$ along the pole.
\label{apndx_2}
\end{appendix}

\end{document}